\documentclass{article}

\usepackage{graphicx}%
\usepackage{multirow}%
\usepackage{amsmath,amssymb,amsfonts}%
\usepackage{amsthm}%
\usepackage{mathrsfs}%
\usepackage{xcolor}%
\usepackage{textcomp}%
\usepackage{manyfoot}%
\usepackage{booktabs}%
\usepackage{algorithm}%
\usepackage{algorithmicx}%
\usepackage{algpseudocode}%
\usepackage{listings}%
\usepackage{array}
\usepackage{caption}
\usepackage{pdflscape}
\usepackage[numbers]{natbib}
\usepackage{subcaption}
\usepackage{threeparttable}
\usepackage{hyperref}
\usepackage[draft]{fixme}
\fxsetup{inline,nomargin,theme=color}

\usepackage{graphicx} %
\usepackage{booktabs}
\usepackage[margin=1in]{geometry}

\theoremstyle{thmstyleone}%

\theoremstyle{thmstyletwo}%

\theoremstyle{thmstylethree}%

\title{A large language model-based approach to quantifying the effects of social determinants in liver transplant decisions}
\author{Emily Robitschek$^{1,2,3}$, Asal Bastani$^{1}$, Kathryn Horwath$^{1}$, Savyon Sordean$^{1}$, \\Mark J. Pletcher$^{1}$, Jennifer C. Lai$^{1}$, Sergio Galletta$^{2}$, Elliott Ash$^{2}$, \\Jin Ge$^{*1}$, Irene Y. Chen$^{*1,3}$}
\date{$^1$University of California, San Francisco, $^2$ETH Zurich, $^3$University of California, Berkeley}

\begin{document}

\maketitle

\abstract{Patient life circumstances, including social determinants of health (SDOH), shape both health outcomes and care access, contributing to persistent disparities across gender, race, and socioeconomic status. Liver transplantation exemplifies these challenges, requiring complex eligibility and allocation decisions where SDOH directly influence patient evaluation. We developed an large language model (LLM)-driven framework to analyze how broadly defined SDOH---encompassing both traditional social determinants and transplantation-related psychosocial factors---influence patient care trajectories. Using large language models, we extracted 23 SDOH factors related to patient eligibility for liver transplantation from psychosocial evaluation notes. These SDOH ``snapshots'' significantly improve prediction of patient progression through transplantation evaluation stages and help explain liver transplantation decisions including the recommendation based on psychosocial evaluation and the listing of a patient for a liver transplantation. Our analysis helps identify patterns of SDOH prevalence across demographics that help explain racial disparities in liver transplantation decisions. We highlight specific unmet patient needs, which, if addressed, could improve the equity and efficacy of transplant care. While developed for liver transplantation, this systematic approach to analyzing previously unstructured information about patient circumstances and clinical decision-making could inform understanding of care decisions and disparities across various medical domains.}

\maketitle

\section{Introduction}\label{sec1}

Healthcare access and outcomes remain fundamentally shaped by social and economic circumstances~\cite{marmot2005social,marmot2012european}, but quantifying these relationships has proven challenging~\cite{truong2020utilization,heidari2023z,wang2021documentation,chen2022clustering,McKinney_Sieniek.2020,Obermeyer_Powers_Vogeli_Mullainathan.2019,guevara2024large,chen2020treating}. Nowhere is this more evident than in liver transplantation (LT), where scarce organs must be allocated based on both medical need and psychosocial stability~\cite{Cleveland_Clinic.2021, Lucey_Furuya_Foley.2023, eCFR.2024, CMS.2021, Duda.2005}. Although the Model for End-Stage Liver Disease (MELD) score provides a standardized measure of medical urgency~\cite{UNOS.2023}, transplant decisions incorporate extensive psychosocial assessments that are often documented in unstructured clinical notes. These assessments capture crucial factors such as history of substance use and social support systems that directly influence transplant eligibility and outcomes~\cite{Kardashian.2022, KARLSEN.2022, ge2023novel,flemming2022association}.

Current transplant evaluation practices mandate documentation of these psychosocial factors~\cite{OPTNetwork.2024a, OPTNetwork.2024b}, but their unstructured nature has hindered large-scale analysis of how they influence care decisions, such as addition of a candidate to the waiting list for transplantation. This limitation is particularly significant given the persistent disparities in transplant access across gender, race, and socioeconomic status~\cite{nephew2021racial, wahid2021review, warren2021racial, moylan2008disparities,mathur2011sex,rosenblatt2021black,mindikoglu2010gender,cullaro2018sex,locke2020quantifying,flemming2022association,ge2023novel}. Understanding how social factors shape transplant decisions requires methods to systematically analyze previously inaccessible information in clinical notes. Recent advances in natural language processing (NLP) and large language models (LLMs) have enabled systematic extraction of social and economic circumstances~\cite{Guevara_Chen.2024,patra2021extracting,lybarger2023leveraging,antoniak2024nlp}, offering a promising approach to address these challenges.

We developed an artificial intelligence (AI) framework that extracts standardized representations of patient circumstances from transplant evaluation notes across 23 dimensions of social determinants using LLMs. Our analysis demonstrates four key findings. First, LLMs can reliably extract social determinants of health (SDOH) factors related to LT decisions as defined by clinicians and social workers. Second, we identify the SDOH factors that vary the most across different patient subgroups and with different policy changes over time. Third, patterns of adverse social factors vary systematically between demographic groups and explain observed disparities in listing decisions. Lastly, these SDOH ``snapshots'' substantially improve the prediction of progression through transplant evaluation when combined with clinical data.  

We present a systematic analysis that demonstrates that previously unstructured information can reveal hidden patterns in how patient circumstances influence transplant decisions, while revealing persistent disparities that warrant attention. The primary focus of our study is to leverage SDOH snapshots to understand liver transplantation decisions by creating standardized representations of patient circumstances at the time of liver transplant evaluation. Our SDOH analysis covers over ten years of LT data from the University of California, San Francisco, a large academic transplant center that conducts over 200 liver transplants per year and serves patients across the Western United States, including California, Nevada, Oregon, Washington, and Hawaii. We quantified the effects of SDOH within subgroups based on race and ethnicity, as well as sex. Our assessment uncovers compelling new insights into how large-scale machine learning (ML) methods can illuminate the impact of SDOH factors on LT decisions, especially in the context of existing health disparities. Because SDOH account for a significant number of modifiable factors related to health outcomes~\cite{hood2016county}, our work creates actionable insights to help clinicians and healthcare professionals address high-impact SDOH factors.

\section{Results}\label{sec2}
\subsection{Datasets and model training}

\begin{figure}[htbp]
    \centering
    \includegraphics[width=0.8\textwidth]{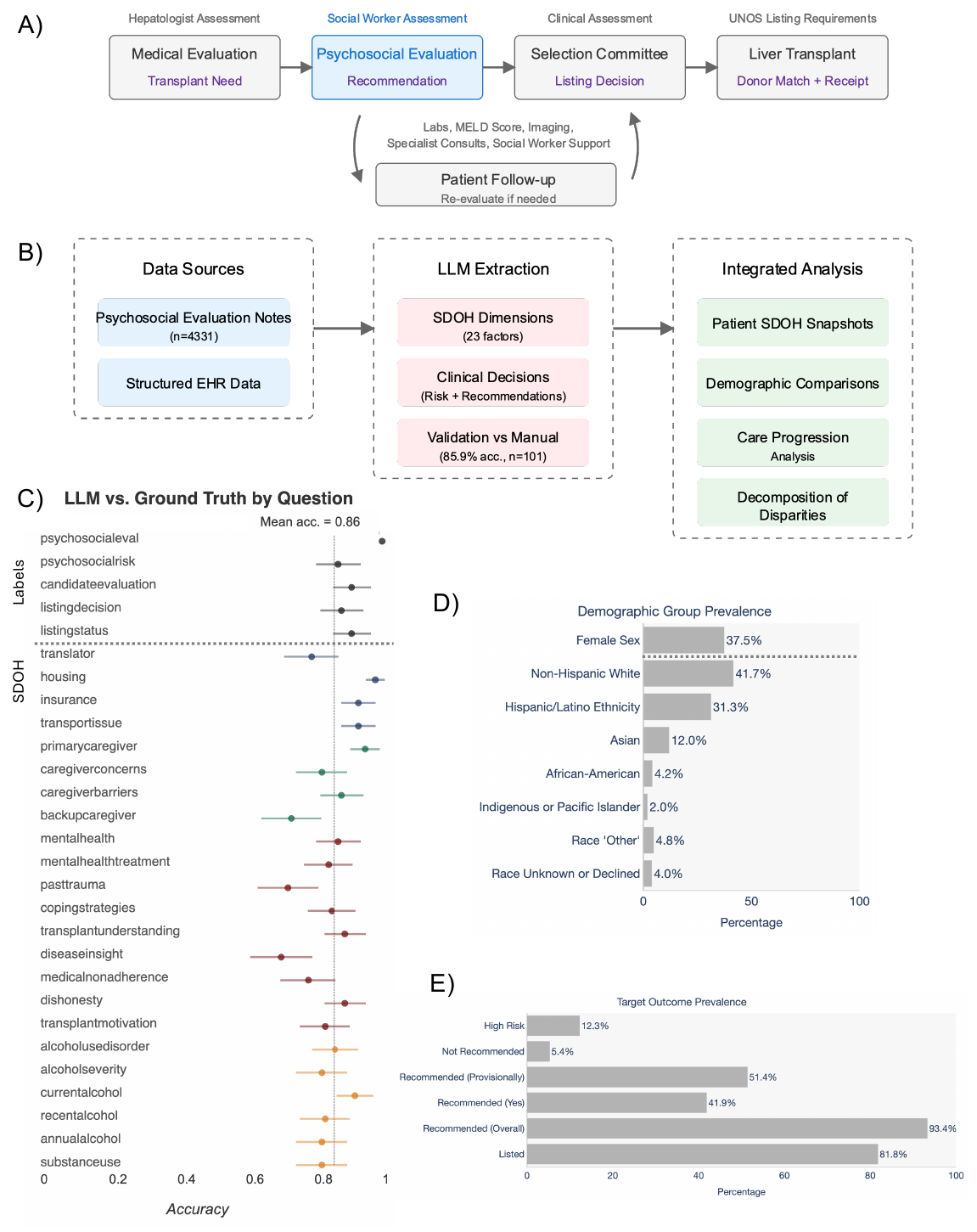}
    \caption{ \textbf{Framework for extracting and analyzing SDOH information from transplant evaluation notes.} a) Schematic overview of liver transplant care journey. Decision outcomes shown in purple. b) Schematic overview of SDOH snapshot creation and analysis pipeline. Clinical notes are processed using LLMs to extract both (i) 23 SDOH dimensions describing patient circumstances* and (ii) clinical decisions/outcomes not captured in structured data (e.g., psychosocial risk assessments, transplant recommendations). These extracted elements are combined with structured clinical and demographic data from the EHR to create comprehensive patient snapshots at evaluation. The integrated data enables (i) comparison of SDOH factor prevalence across demographic groups, (ii) identification of transition points where specific factors impact care progression, and (iii) decomposition analysis of how SDOH patterns and clinical factors explain demographic differences in care access. This approach surfaces both individual-level circumstances and population-level patterns that can guide resource allocation and policy decisions. c) Accuracy of GPT-4-Turbo-128k vs. ground truth annotations (n=101) for 28 questions, including 23 SDOH-related dimensions. d) Demographic composition of the study cohort (n=3,704). e) Prevalence of key clinical outcomes, including psychosocial recommendation status (Rec) and liver transplant (LT) listing rates. *SDOH colored by related theme (yellow=`Substance Use'; green=`Social Support'; blue=`Access', and red=`Psychological')}
        \label{fig:figure1}
    \end{figure}

We conducted a retrospective, longitudinal analysis using deidentified electronic medical record (EMR) data collected at the University of California, San Francisco. Our patient cohort includes 4,331 adult patients evaluated for liver transplantation (LT) between 2012 and 2023 and extracted psychosocial evaluation notes from these patients. The final cohort (n=3,704) included patients with complete demographic and clinical data. See the study diagram in Figure~\ref{fig:figure1} and Figure \ref{fig:cohort_diagram} for the cohort selection diagram.

Our evaluated cohort is demographically diverse (Figure \ref{fig:figure1}D), and the majority of patients have the full set of structured clinical data (n=3,704). The distribution of patient race and ethnicity within the cohort is 42\% Non-Hispanic White, 31\% Hispanic or Latino, 13\% Asian, 4\% Black or African American, 2\% Indigenous and Pacific Islanders, 5\% Other, and 3\% ``Unknown or Declined'' (unknown race or undisclosed race), with a gender distribution of 37.5\% female and 62.5\% male.  Within this cohort, 41\% of the patients have a diagnosis of hepatocellular carcinoma (HCC), with higher prevalence for Asian patients (56.8\%), and lower prevalence than the average for female and Hispanic / Latino patients having (30.1\% and 37.6\% respectively). Controlling for HCC exceptions, there are no statistically significant differences in liver disease severity between demographics at a population level, as estimated by MELD scores (Figure \ref{fig:liver_disease_comparison}).

We analyze two clinical decisions in the liver transplant care journey (Figure~\ref{fig:figure1}A). Each potential LT patient is assessed for psychosocial factors, which is documented in the psychosocial evaluation. Our first clinical decision is the recommendation of the psychosocial evaluation. After a psychosocial recommendation is given, the selection committee combines the recommendation with other clinical assessments to decide whether or not to list this LT patient with the United Network of Organ Sharing (UNOS). Our second clinical decision is the LT listing decision. Due to the many complexities in the transplant matching process, we focused on the immediate effects of the decision of the LT panel compared to the more complex outcomes related to the actual completion of the transplant. The detailed prevalence rates for transplant listing by demographic group for all evaluated patients  are shown in Table~\ref{table:demographics_by_list_LT}. 

To extract SDOH factors, we employed a HIPAA-compliant LLM (gpt-4-turbo-128k). To assess the benefits of SDOH factors on predictive performance, we trained an Extreme Gradient Boosting (XGBoost) model. To understand and quantify the effects of SDOH on the LT decision-making process, we examined whether the addition of SDOH factors would improve the performance of a predictive model trained to predict LT outcomes. Specifically, for each patient evaluated for LT, we make two predictions: whether they will be recommended for LT based off of the psychosocial evaluation and whether they will be listed LT. These two prediction models are developed using a combination of clinical, demographic, and extracted SDOH factors. We examined the interpretability of the predictive model using SHAP values~\cite{Lundberg_Lee.2017}, which indicate the features that most impact the model predictions and the direction and strength of that impact.

\subsection{LLMs can extract liver transplant SDOH factors}

Building on recent work that demonstrated that LLMs can extract SDOH factors~\cite{guevara2024large}, we developed a large-scale SDOH extraction pipeline that could amplify the domain expertise of LT specialists. 
We defined 23 SDOH categories based on the recent literature and hospital policies~\cite{Kardashian.2022, KARLSEN.2022} in close collaboration with licensed clinical social workers and a transplant clinician. These categories include any history of substance use, access factors for the patient, social support, and mental health factors (Figure~\ref{fig:figure1}C). The text description shown shows a few word summary of the factor derived from each question in the note query (see full table of questions in Supplementary Table \ref{tab:llm_questions_categories}). Using our HIPAA-compliant LLM, we extracted these dimensions from psychosocial evaluation notes of patients considered for LT. The length of the notes ranged from 221-4,972 tokens (mean: 1,592, SD: 507). We created a ``SDOH snapshot'' for each patient, capturing key factors that can influence LT outcomes. The extraction accuracy was validated against 101 expert annotations from licensed clinical social workers and a transplant clinician.

When validated against manual annotations in a random subset (n=101), our approach achieved an average accuracy of 0.859 across all SDOH categories (95\% CI: 0.846-0.872). The accuracy for the individual SDOH categories ranged from 0.70 (95\% CI: 0.61-0.79) for the disease insight of the patient to 0.98 (95\% CI: 0.95-1.00) for housing instability (Figure \ref{fig:figure1}C). Our results demonstrate that LLMs can reliably extract LT SDOH snapshots comparable to human subject matter experts.

\subsection{SDOH factors reveal differing patterns across patients}
We computed the differences in SDOH snapshots across patient subgroups. Because not all patient data contains complete clinical information, this missingness and censorship can affect downstream analysis and models~\cite{chen2022clustering}. As a result, we studied patients with complete clinical information and patients with incomplete clinical information separately. Among patients with complete clinical data (n=3,704), we observed significant demographic variations in SDOH factors (Figure \ref{fig:figure5}B). We focused initially on this subset to be able to control for the contribution of clinical factors and disease status when modeling SDOH impacts. Asian patients consistently demonstrated lower rates across multiple psychosocial domains compared to the mean baseline, including severe alcohol use (-68.9\%) and mental health treatment (-73.0\%). Gender analysis revealed female patients were 38.2\% more likely to have no identified primary caregiver. Female patients also had higher rates of mental health issues (+31.6\%) and ongoing treatment (+36.0\%), and of past trauma (+36.6\%), while showing lower rates of severe alcohol use (-35.8\%). Hispanic/Latino patients exhibited higher rates of mental health issues (+16.8\%) but lower reported rates of mental health treatment (-14.\%) - a possible gap of 31.5\%. Indigenous/Pacific Islander patients showed higher rates in transportation access needs (+138.2\%) and history of medical non-adherence (+85.6\%). Non-Hispanic White patients showed higher rates of substance use (+22.1\%) while experiencing lower rates of transportation issues (-23.9\%). `Other' race patients showed a higher prevalence of history of medical non-adherence (+56.5\%) and translator need (+75.5\%). Patient with unknown race or that declined to specify their race to UCSF tended to have higher rates of severe alcohol use (+24.1\%), ongoing mental health treatment (+43.3\%) and a +55.7\% lower rate of backup caregivers identified. 

Some patients in our study lacked complete clinical data (n=548), potentially due to greater vulnerability or early loss to follow-up. This group revealed generally higher adverse SDOH prevalence (Supplementary Figure~\ref{fig:dem_var_missing_meta}). To avoid overlooking important variations potentially associated with these more vulnerable patients, we then examined the larger cohort with demographic data (n=4,243). While our previous findings of SDOH prevalence across demographics remained robust, several additional associations emerged. African-American patients showed higher rates of low transplant process knowledge (+70\%) and non-alcohol substance use (+29.8\%). Patients with undisclosed race demonstrated increased rates across multiple dimensions, including suspected dishonesty (+40.4\%), poor coping skills (+87.2\%), and were more likely to have caregivers about whom social workers expressed concerns regarding adequate support provision (+43.7\%). The expanded analysis also revealed that Indigenous/Pacific Islander patients showed higher rates of severe alcohol use history (+29.0\%). Female patients demonstrated higher prevalence of poor coping skills (+21.2\%) while maintaining all previously described SDOH factor associations. Transportation issues were significantly higher for patients of other (+74.6\%) and undisclosed race (+67.2\%).

Finally, focusing specifically on the subset of patients without full information (n=548) revealed distinct patterns of SDOH prevalence. Social workers were more likely (+52.1\%) to identify potential barriers to the primary caregiver's ability to provide adequate support for patients of undisclosed race. Asian patients in this subset maintained lower prevalence of substance use and psychosocial risk factors compared to other demographics, but showed notably higher translator needs than the overall cohort (+252.7\%). Female patients in this group continued to show lower rates of alcohol use disorder (-19.7\%) and prior severe alcohol use (-22.6\%), while maintaining higher prevalence of past trauma (+43.0\%), ongoing mental health challenges (+19.5\%), and mental health treatment (+31.3\%). 

\subsection{SDOH factors reveal temporal shifts}
Temporal analysis revealed notable shifts in both demographics and psychosocial health factors over the study period, including an increase in proportion of Latino patients from 22\% in 2012 to 43\% in 2023 (Figure \ref{fig:year_dem}) and rising rates of recent alcohol use (18\% in 2012 to 28\% in 2023) (Figure \ref{fig:year_substance}). Patients requiring translation services steadily increased from 14.5\% to 20.8\% over the same time period (Figure \ref{fig:year_access}). These increases potentially reflect transplant policy changes at UCSF and epidemiological shifts in liver disease burden. The observed temporal increase in documented mental health factors like trauma (11\% in 2012 to 18\% in 2023) and mental health challenges (24\% in 2012 to 36\% in 2023) may reflect evolving screening practices and reduced stigma rather than true trends in prevalence. Leveraging the snapshots in this way we can see both recent `shocks' (the jump in the prevalence of reported ongoing mental health issues from 2022 to 2023), and longer term trends in the case of the reporting of past trauma, translator requirements, and the increase in the proportion of Hispanic/Latino patients evaluated for transplant at UCSF.  

\subsection{SDOH factors explain racial disparities}
Racial disparities in LT are well-documented, but the mechanisms underlying these differences remain poorly understood. We hypothesized that SDOH factors might explain a substantial portion of these disparities. In our dataset, Asian patients showed significantly higher listing rates while patients with unknown or undisclosed race showed markedly lower rates compared to the patient average, making these populations particularly informative to understanding disparities (Figure \ref{fig:figure6}A). To rigorously quantify the how much of these differences could be explained by measurable factors, including SDOH, we employed the Blinder-Oaxaca decomposition method--a statistical technique that quantifies how much of an outcome gap between groups can be attributed to differences in measured characteristics versus other unexplained factors (Figure \ref{fig:figure5}C). For Asian patients, SDOH factors in isolation explained 42.6\% of listing outcome gaps - more than measures of liver health, which explained 36.8\%. Combined features explained 94.6\% of variance in Asian patient listing outcomes. However, for patients with unreported race that is either unknown or undisclosed, only 10.8\% of listing outcome gaps were explained by combined features (8.6\% liver health, 3.6\% SDOH).

While the Blinder-Oaxaca analysis revealed how much of the racial disparities could be explained by measured factors, we next sought to understand which specific SDOH factors had the strongest influence on listing decisions. To investigate this, we identify the SDOH factors most negatively associated with recommendation and listing (Figure~\ref{fig:figure6}B), with listed patients less likely to have low transplant motivation (-55.6\%), current alcohol use (-28.5\%), poor coping skills (-28.5\%), and unstable housing (-23.6\%). They are also less likely to lack a primary caregiver (-24.4\%). To further examine the relative individual contributions of clinical, demographic, and SDOH factors to listing likelihood, we carry out regression analysis (Figure~\ref{fig:figure6}C). We find that when controlling for clinical factors and SDOH factors, some racial differences were no longer statistically significant. However, other racial information such as having an unknown or undisclosed race race greatly decreased the rate of LT listing (-0.13, p$<$0.001), further indicating possible disparities unexplained by SDOH factors for these patients. As expected, clinical factors such as hepatocellular carcinoma (HCC) (0.12, p$<$0.001) and MELD score (0.03, p$<$0.001) have significant positive impact on transplant listing; however, some SDOH factors such as good disease insight (0.08, p$<$0.001) also have a positive impact on listing above and beyond a standard deviation change in MELD score. SDOH factors with negative effects on LT listing outcome include current alcohol use (-0.14, p$<$0.05), concerns with the primary caregiver (-0.08, p$<$0.01), past trauma (-0.07, p$<$0.01), ongoing unmanaged mental health challenges (-0.06, p$<$0.01), and other non-alcohol substance use (-0.05, p$<$0.01). Our analysis reveals important patterns in how adverse social determinants of health cluster together (Figure~\ref{fig:sdoh_cooccur}). Patients with alcohol use disorder or severe substance use histories frequently face multiple concurrent social challenges. The co-occurrence analysis particularly highlights the relationship between active mental health challenges and other adverse factors, including past trauma, limited coping skills, and housing instability. We also observe clustering among factors related to medical understanding - specifically, patients with limited disease insight often also demonstrate poor coping skills and insufficient transplant knowledge (Figure~\ref{fig:sdoh_cooccur}). 

\subsection{SDOH factors improve prediction for LT psychosocial recommendation and listing}
Recognizing that comparisons across demographic groups do not give a complete picture of the influence of SDOH factors on individual patient outcomes, we leverage non-linear predictive modeling to provide a more individualized understanding how SDOH factors can influence prediction of LT decisions.
We demonstrated that integration of LLM-derived SDOH features with demographic information and clinical covariates including MELD score, HCC status and BMI significantly improved outcome prediction. For psychosocial recommendation, the area under the receiver-operator curve (AUROC) increased 77.3\% from 0.494 (95\% CI: 0.413-0.585) to 0.876 (95\% CI: 0.838-0.915) (Figure \ref{fig:figure2}). For eventual successful listing, AUROC improved 16.4\% from 0.616 (95\% CI: 0.566-0.666) to 0.717 (95\% CI: 0.670-0.762) (Figure \ref{fig:figure3}). 

Among patient recommended based on their psychosocial evaluations, models predicting successful listing achieved an AUROC of 0.589 (95\% CI: 0.534-0.643) using clinical features alone (Figure 4). Including SDOH factors increased AUROC to 0.680 (95\% CI: 0.628-0.732)--a 15.5\% improvement. SDOH-only models achieved an AUROC of 0.641 (95\% CI: 0.586-0.696), exceeding clinical-feature models and suggesting SDOH factors may better predict listing outcomes than traditional clinical metrics including measures of liver health (i.e., MELD score, HCC status) and other characteristics known to influence likelihood of transplant (e.g. BMI and age). These results indicate SDOH impact extends beyond initial recommendations and likely captures patient characteristics distinct from clinical measures.

Shapley Additive exPlanations (SHAP) analysis was carried out to help identify key predictors across the transplant process and assess the relative importance of clinical and SDOH features (Figure \ref{fig:figure4}). For psychosocial recommendation, top predictors were lack of medical non-adherence, lack of caregiver concerns or dishonesty, good coping skills, and absence of non-alcohol substance use. Successful listing was positively associated with HCC diagnosis, MELD score, a lack of caregiver concerns, no alcohol use in the past year, and good disease insight, while substance use, poor coping skills, ongoing mental health challenges, and higher age had negative impact. 

We benchmark how LLM-derived SDOH factors compare to NLP baseline models such as Bag-of-Words (BOW) and clinical Text Analysis and Knowledge Extraction System (cTAKES) concepts~\cite{savova2010mayo}. We assessed two binary outcomes: psychosocial recommendation (93\% base rate) and eventual successful listing (81\% base rate) (Figure \ref{fig:figure1}D). Models using cTAKES features performed the worst (e.g., psychosocial recommendation AUROC of 0.52; 95\% CI 0.44-0.60). Models using LLM-derived and BOW features performed higher (Supplementary Figure \ref{fig:rec_curves_text_only}). The model using BOW features outperformed the model using LLM features: for psychosocial recommendation, the model using BOW features has an AUROC of 0.91 (95\% CI: 0.86-0.95) whereas the LLM features has AUROC of 0.87 with 95\% CI: 0.82-0.90. However, we note that the LLM features are far more interpretable and seem to avoid the label leakage that likely contributes to the performance of the models based on BOW features (Supplementary Figure~\ref{fig:shap_text_only}).

\section{Discussion}\label{sec12}

\begin{figure}[htbp]
    \centering
    \includegraphics[width=0.9\textwidth]{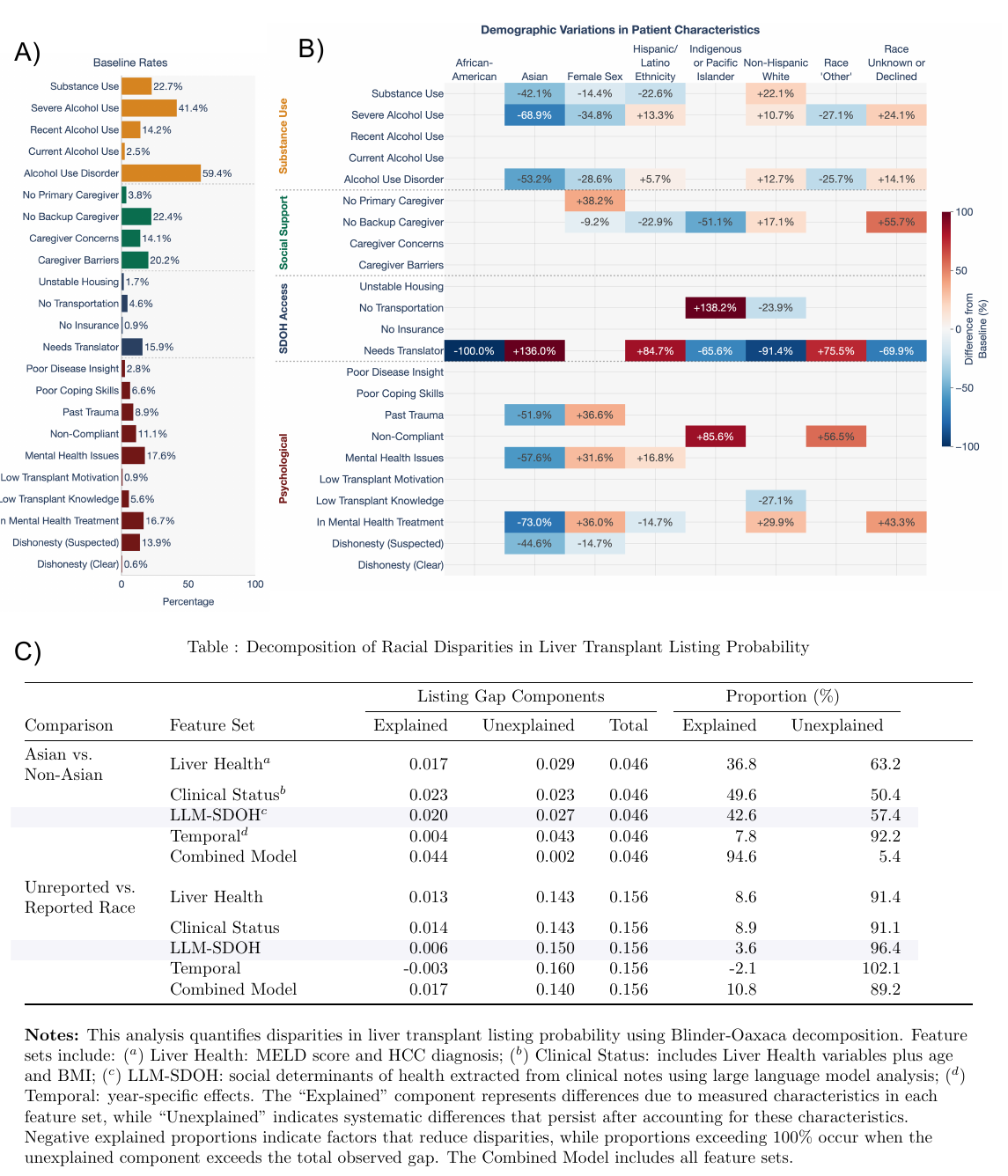}
    \caption{\textbf{Analysis of demographic disparities in liver transplant listing rates.} a) Baseline prevalence rates for psychosocial and substance use factors identified in clinical notes. b) Heat map showing statistically significant differences in factor prevalence across demographic groups (two-proportion z-tests, p $<$ 0.05, FDR-corrected), expressed as percentage point differences from baseline; colored boxes represent statistically significant differences from patient average; blue indicates lower rates, red indicates higher rates. c) Blinder-Oaxaca decomposition analysis quantifying explained and unexplained components of listing probability disparities, showing independent contributions of liver health metrics, SDOH features, and temporal effects.}
        \label{fig:figure5}
    \end{figure}

\begin{figure}[htbp]
    \centering
    \includegraphics[width=0.9\textwidth]{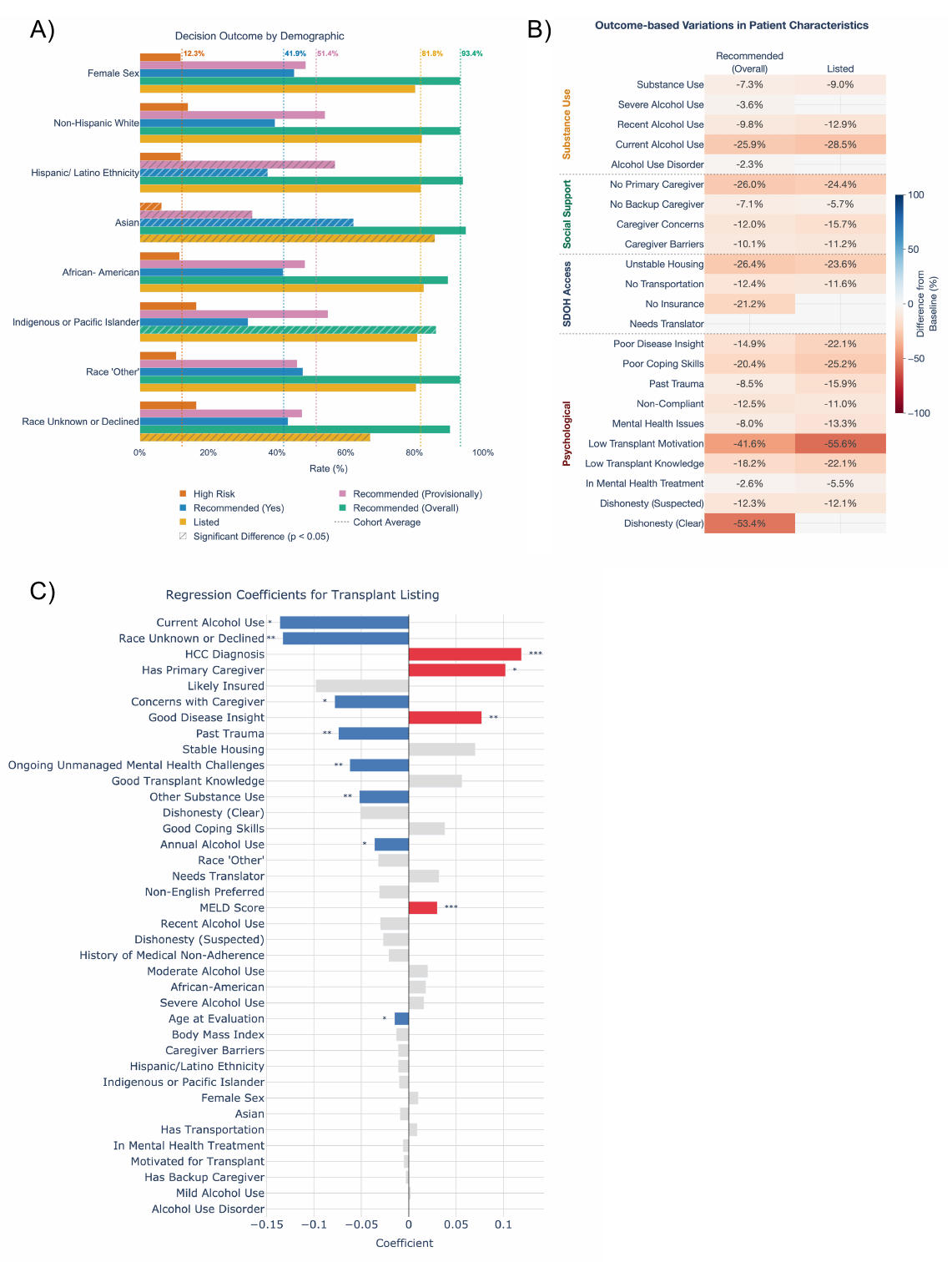}
    \caption{\textbf{Demographic and SDOH variation across liver transplant outcomes.} 
    a) Percentage of patients reaching each evaluation milestone$^{\dag}$  stratified by demographic group, showing progression from initial psychosocial risk assessment through listing. Striped bars indicate significant differences from overall cohort means (FDR-corrected two-proportion z-tests). b) Heat map showing significant differences in SDOH factor prevalence between patients who did versus did not achieve each outcome (two-proportion z-tests, p $<$ 0.05, FDR-corrected); blue indicates higher rates, red indicates lower rates, blank cells indicate non-significant differences. c) OLS regression coefficients with LLM-derived, clinical, and demographic features. Significant coefficients marked (* p$<$0.05, ** p$<$0.01, *** p$<$0.001) and colored based on whether they have a positive (red) or negative (blue) impact on listing. $^{\dag}$Note on outcome classifications: ``Recommended (Yes)'' refers only to patients receiving unconditional recommendations, while ``Recommended (Provisionally)'' is a separate group. These two recommendation types are mutually exclusive. The ``Overall'' group includes both provisionally and unconditionally recommended patients. All other outcomes can co-occur.}
        \label{fig:figure6}
    \end{figure}

\begin{figure}[htbp]
    \centering
    \includegraphics[width=0.9\textwidth]{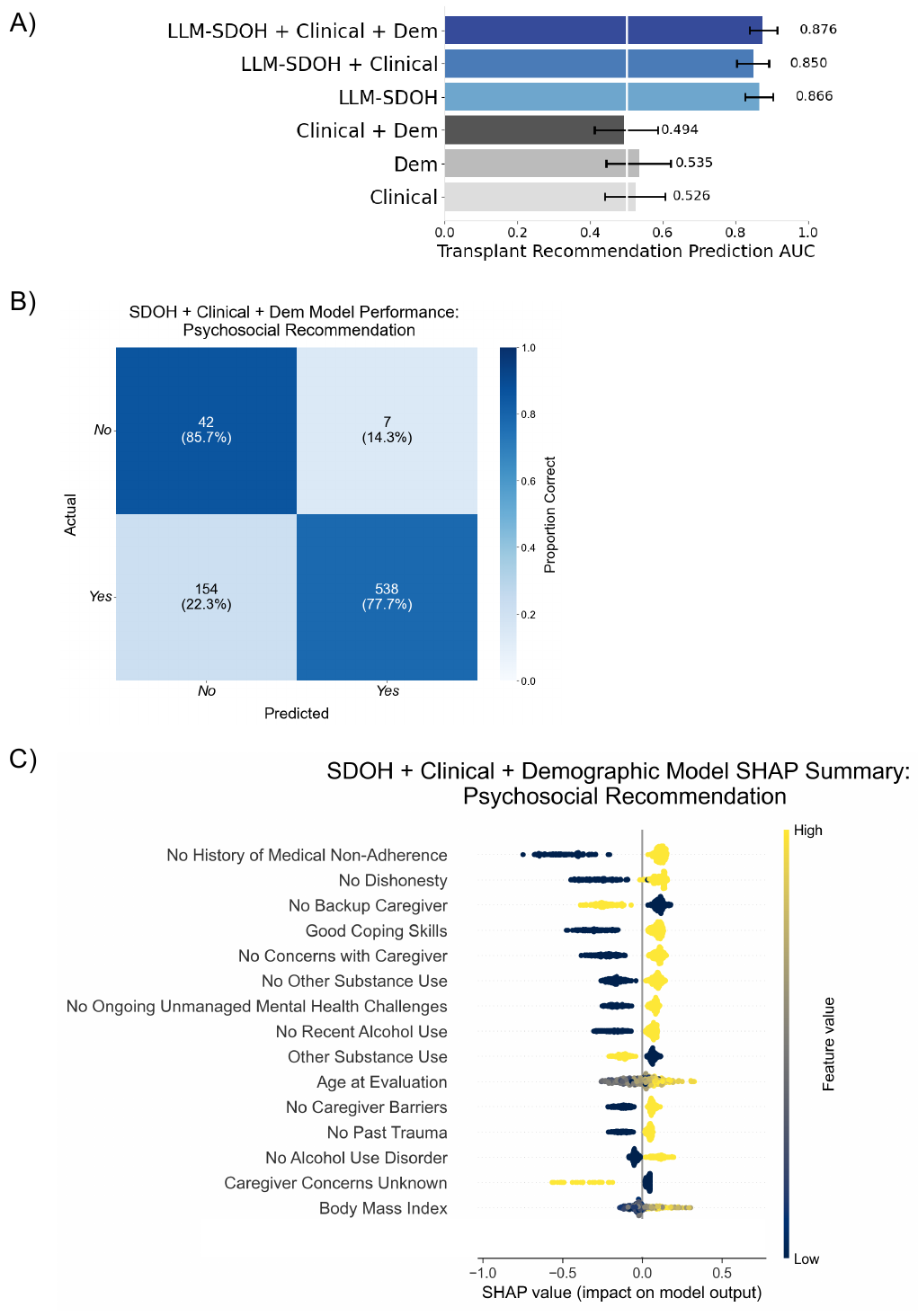}
    \caption{
    \textbf{Model performance and feature analysis for psychosocial recommendation prediction.} a) Comparison of average AUROC (w. 95\% CI) across six combinations of  clinical, demographic, and LLM-derived feature sets. Feature sets including LLM-derived features shown in blue. b) Confusion matrix for the LLM-SDOH + Clinical + Demographic combined feature model with normalized percentages over true values (rows). c) SHAP (SHapley Additive exPlanations) values for the top 15 features for the model with all feature sets.}
        \label{fig:figure2}
    \end{figure}

\begin{figure}[htbp]
    \centering
    \includegraphics[width=0.95\textwidth]{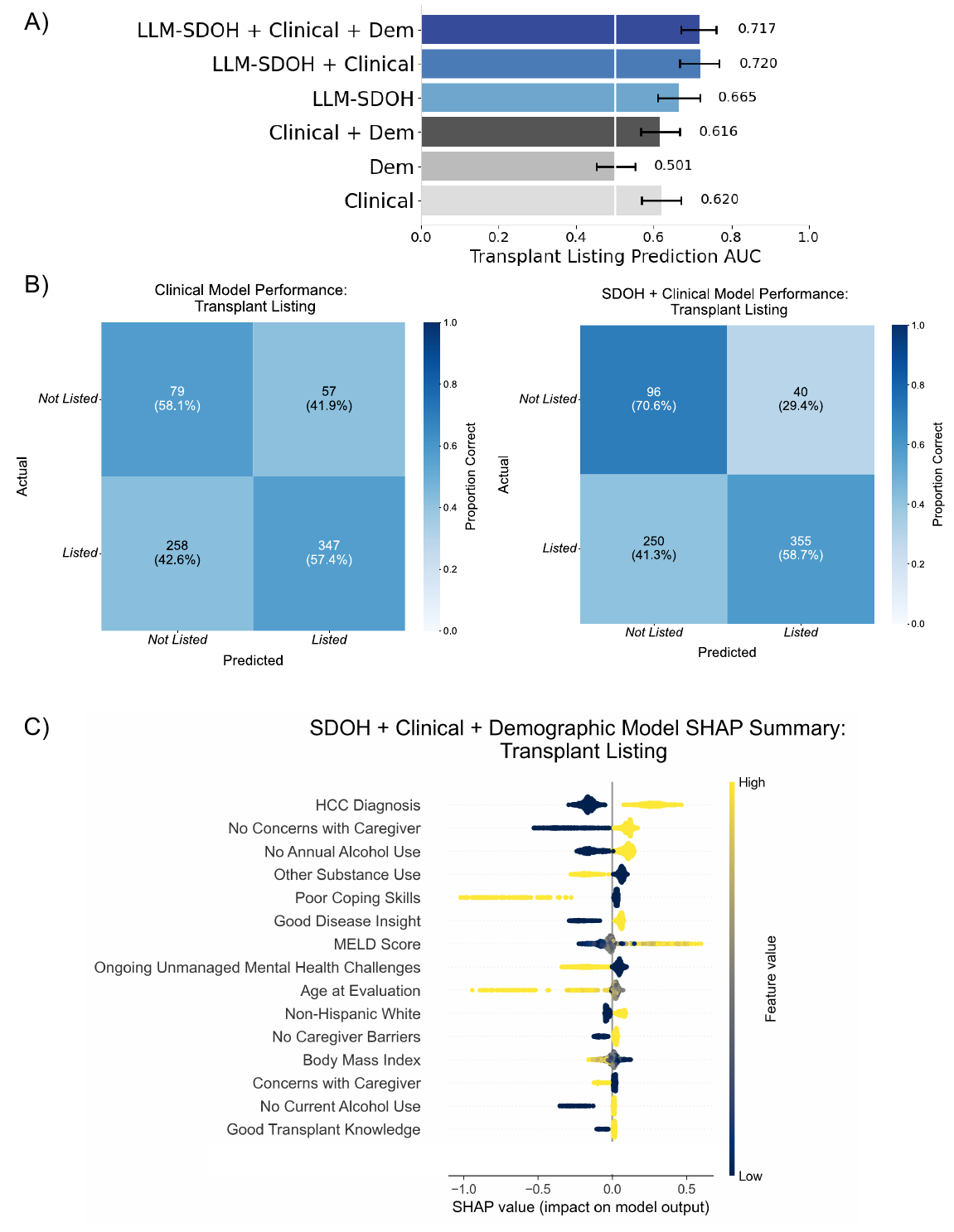}
    \caption{\textbf{Model performance and feature analysis for liver transplant listing prediction.} a) Comparison of average AUROC (w. 95\% CI) across six combinations of  clinical, demographic, and LLM-derived feature sets. Feature sets including LLM-derived features shown in blue. b) Confusion matrix for the Clinical (left) and Clinical + LLM-SDOH combined feature model (right) with normalized percentages over true values (rows). c) SHAP (SHapley Additive exPlanations) values for the top 15 features for the model with all feature sets.}
        \label{fig:figure3}
    \end{figure}

\begin{figure}[htbp]
    \centering
    \includegraphics[width=0.9\textwidth]{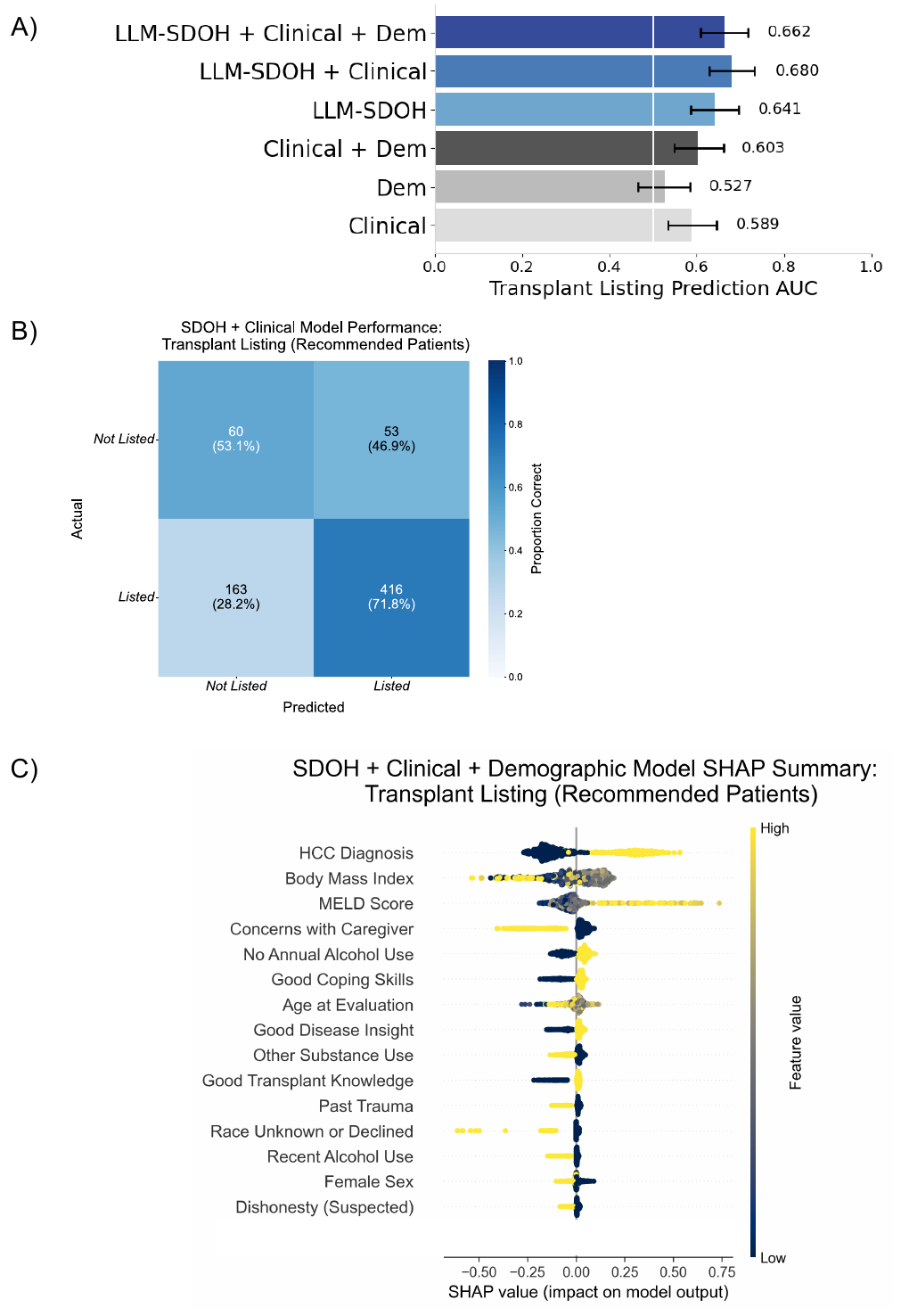}
    \caption{\textbf{Model performance and feature analysis for liver transplant listing prediction in patients with psychosocial recommendations.} a) Comparison of average AUROC (w. 95\% CI) across six combinations of  clinical, demographic, and LLM-derived feature sets. Feature sets including LLM-derived features shown in blue. b) Confusion matrix for the Clinical + LLM-SDOH combined feature model with normalized percentages over true values (rows). c) SHAP (SHapley Additive exPlanations) values for the top 15 features for the model with all feature sets.}
        \label{fig:figure4}
    \end{figure}

Our findings demonstrate that systematic extraction of SDOH information from clinical notes improves understanding of liver transplant access and outcomes. Through our analysis, we show that LLMs can reliably extract patient circumstances from unstructured documentation, achieving 0.70-0.98\% accuracy across multiple SDOH dimensions (Figure~\ref{fig:figure1}). This capability proves valuable given our finding that SDOH-only models outperform clinical feature models in predicting successful listing, indicating these factors capture unique and non-redundant information about patient progression through the transplant evaluation process (Figure~\ref{fig:figure3}, ~\ref{fig:figure4}). Our analysis identifies key SDOH factors contributing to both LT decisions including current or recent alcohol use, lack of social support (e.g. primary caregiver), unstable housing and medical-literacy-associated factors such as disease insight, coping skills, transplant knowledge (Figures~\ref{fig:figure6},~\ref{fig:figure2},~\ref{fig:figure3}, and~\ref{fig:figure4}). These dimensions also help explain racial disparities in our cohort, and highlight remaining unexplained disparities for follow up (Figure~\ref{fig:figure5}). We emphasize that the intent of our modeling is to understand the effects of SDOH on liver transplant outcomes and decisions, not to provide an AI tool to reproduce those decisions. Beyond our quantitative results, our work has several generalizable insights and larger implications for the medical community. 

First, our work directly ties machine learning approaches to SDOH factors relevant to LT decision making that can provide guidance for healthcare practitioners. As larger medical datasets and LLMs provide potentially more value to medical researchers~\cite{guevara2024large, chen2020treating, shen2024data}, it is very likely that other liver transplantation centers may find slightly different patterns and biases depending on their patient population and policy decisions; however, our framework enables the understanding of specific SDOH factors that significantly affect outcomes at different stages. 

The lack of social support emerges as a major barrier in our UCSF cohort. Specifically, the absence of primary caregivers is associated with reduced progression through psychosocial recommendation and LT listing, with caregiver-related factors emerging as significant predictors of outcomes across models. Medical literacy and psychological resilience represent another critical barrier. Key predictive factors include disease insight, coping skills, history of medical non-adherence, and ongoing mental health challenges. These findings point to several potential interventions: structured caregiver support programs including financial compensation, expanded psychoeducation through multilingual recorded sessions, and enhanced peer support through transplant mentors and support groups. Our temporal analysis further reveals increasing documentation of translator needs, alcohol use, and mental health concerns, suggesting growing demands for integrated translation, substance use, and mental health services. 

The demographic patterns in our analysis point to specific opportunities for intervention. In our patient cohort, Indigenous/Pacific Islander patients face higher rates of transportation barriers and lower initial recommendation rates, suggesting a need for enhanced support services during early evaluation stages. Hispanic/Latino patients show higher rates of mental health issues but lower rates of ongoing treatment, while being more likely to receive provisional recommendations. This pattern indicates potential benefits from integrated, culturally-sensitive mental health support combined with transplant evaluation follow-up. These findings become apparent when examining our complete patient cohort, showing how analyses restricted to patients with complete clinical data may underestimate both the prevalence and impact of certain SDOH factors. In order to model the range of interventions available, our binary classification task of recommendations could be expanded to make distinctions between provisional and full recommendations for care.

Second, our framework of developing standardized SDOH representation to model the effects of social and economic circumstances could be used to model a wide range of medical applications---including other organ transplantation decisions, maternal health outcomes, and care for chronic conditions. The rapid proliferation of LLMs used on clinical data~\cite{clusmann2023future} has demonstrated tremendous potential to improve health equity. Although researchers have raised concerns about the potential for ML models and particularly LLMs to amplify existing disparities~\cite{Chen_Pierson_Rose_Joshi_Ferryman_Ghassemi.2021}, we believe that careful construction and validation of models could enable a new wave of understanding the interplay of both human and ML bias to improve health outcomes.

Third, we highlight that our LLM extraction of SDOH factors relies on an assumed ground truth when validating SDOH factors. Our analysis implicitly sets social worker labels as the gold standard; however, more scrupulous analysis of automated labeling by LLMs may be warranted, especially since automated NLP labeling techniques may skew downstream ML results for different patient subpopulations. In order to better understand in the full LT psychosocial evaluation process, the transcript of interviews could be studied, potentially with the help of automated NLP tools.

Lastly, although EHR data and newly unlocked unstructured notes can shed more insight into the process, we must be vigilant for documentation biases that can magnify the ``streetlight effect'' --- the tendency to search for answers only where data is easily available ~\cite{hoelzemann2024streetlight}. While SDOH factors explain a substantial portion of listing outcome gaps for Asian patients, significant unexplained differences persist for other groups, particularly patients with undisclosed race. This variation in explanatory power points to a key analytical challenge: those most vulnerable to early loss from the transplant evaluation process often lack complete clinical data, potentially excluding them from traditional analyses. The forces that drive health outcomes extend far beyond clinical encounters, and while we expand analyzable information beyond traditional structured clinical data, we remain limited by what providers choose to document. Despite these challenges, we believe that as machine learning technologies advance and medical datasets increase in size and scope, we have an opportunity to enable a deeper examination of patient life circumstances.

\section{Figures}\label{figures_temp}

\clearpage
\section{Methods}\label{sec11}

\subsection{Data and Preprocessing}
We analyzed psychosocial evaluation notes from 4,331 adult patients evaluated for liver transplantation (LT) at a large academic medical center between 2012 and 2023. The final cohort (n=3,704) included patients with complete demographic and clinical data. The cohort's race-ethnicity distribution was 42\% Non-Hispanic White, 31\% Hispanic or Latino, 13\% Asian, 4\% Black or African American, 2\% Indigenous and Pacific Islanders, 5\% Other, and 3\% Unknown or Declined, with a gender distribution of 37.46\% female and 62.54\% male. Numerical variables were normalized using scikit-learn's StandardScaler \citep{scikit-learn}. Categorical variables were one-hot encoded. For Bag-of-Words (BOW) baseline comparison models, we used NLTK \citep{bird2009natural} for text preprocessing and scikit-learn for feature extraction and selection.

\subsection{SDOH Definition and Extraction}
We defined 23 Social Determinants of Health (SDOH) categories based on recent literature \citep{Kardashian.2022, KARLSEN.2022} and hospital policies. These categories included substance use history, patient access factors, social support, and mental health factors. The categorization was developed in close collaboration with licensed clinical social workers and a transplant clinician.
We employed a privacy-preserving version of GPT-4-Turbo-128k to survey these dimensions from clinical notes, creating a ``SDOH snapshot'' for each patient, capturing key factors that may influence LT outcomes. Extraction accuracy was validated against 101 expert annotations.

\subsection{Model Development}
XGBoost \citep{Chen_Guestrin.2016} models were developed to predict two key outcomes: psychosocial recommendation and  eventual successful listing. We used 80\% (n=2,963) of data for training and hyperparameter tuning and 20\% (n=741) for testing, with stratification by outcome. Models were created using: (1) Clinical covariates only, (2) Clinical covariates + LLM-derived SDOH features, and (3) Clinical covariates + SDOH features + demographic factors. Downsampling of the majority class was performed using RandomUnderSampler \citep{imbal_learn_2017}. Hyperparameter tuning used grid search with 5-fold cross-validation, exploring: max\_depth [3, 6, 9], learning\_rate [0.01, 0.1, 0.2], n\_estimators [100, 300, 500], subsample and colsample\_bytree [0.7, 0.8, 0.9], and gamma [0, 0.1, 0.2]. Ordinary Least Squares (OLS) models were created with the linear model function from the statsmodels python package \citep{seabold2010statsmodels} with `HCV3' robust standard errors.

\subsection{Evaluation}
Model performance was evaluated using area under the receiver-operator curve (AUROC), sensitivity, and specificity on the held-out test set. SHAP values were used to interpret feature importance \citep{Lundberg_Lee.2017}. To further analyze differences in outcomes across demographic groups, we employed linear probability models and Blinder-Oaxaca decomposition \citep{Oaxaca.1973, Rahimi_Hashemi_Nazari.2021, Fortin_Lemieux_Firpo.2011}, both implemented with the python statsmodels package \citep{seabold2010statsmodels}.

\subsection{Statistical Analysis of SDOH by Demographic}
We conducted systematic analyses of psychosocial and clinical factors across demographic groups for the set of patients with all clinical features including admissions information (n=3695) using a structured statistical approach implemented in Python. For each factor, we calculated baseline prevalence rates and demographic-specific variations using two-proportion z-tests with 95\% confidence intervals, implemented through scipy.stats \citep{SciPy-NMeth.2020}. Statistical testing employed the chi2\_contingency and norm.cdf functions from scipy.stats for chi-square tests and z-score calculations, respectively. Multiple comparison adjustment was performed using the multipletests function from statsmodels.stats.multitest \citep{seabold2010statsmodels} with the Benjamini-Hochberg procedure to control false discovery rate across demographic comparisons. 
All statistical tests were conducted with $\alpha$=0.05, and results were stratified by factor domains (Social Support, Access, Psychological, and Substance Use) to enable domain-specific evaluation of demographic patterns. Temporal trends of prevalence across SDOH factors and demographics were visualized with line graphs.

\subsection{SDOH Co-occurrence}
Co-occurrence patterns between adverse SDOH factors were analyzed using a normalized matrix approach. For each pair of binary factors $i$ and $j$, we calculated the percentage of cases where factor $j$ was present given the presence of factor $i$, yielding an asymmetric co-occurrence matrix. The computation was performed using matrix multiplication of binary indicators, with normalization by factor prevalence to obtain conditional percentages. Each cell $(i,j)$ represents the percentage of patients with factor $j$ among those who had factor $i$, calculated as $(n_{ij}/n_i)\times100$, where $n_{ij}$ is the count of patients with both factors and $n_i$ is the count with factor $i$. The diagonal represents 100\% by definition. Results were visualized as a heatmap to highlight co-occurrence patterns, revealing potential compound vulnerabilities in the patient population.

\clearpage
\bibliographystyle{unsrt}
\bibliography{bibliography}

\appendix

\section{Supplementary Figures}\label{secB1}

\begin{figure*}[htbp]
    \centering
        \centering
        \includegraphics[width=1.0\textwidth]{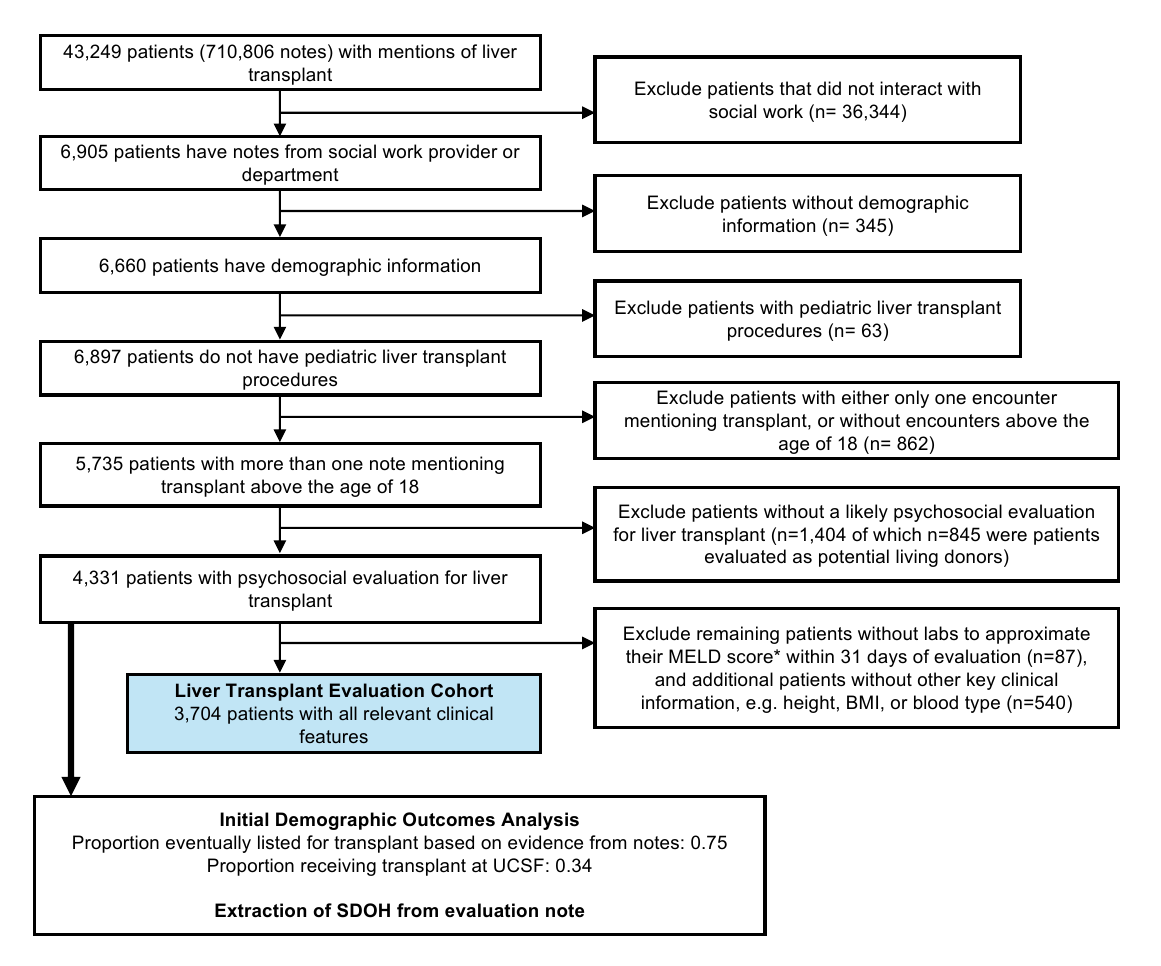}
        \caption{ \textbf{Cohort Selection Process for Liver Transplant Evaluation Study}. This flowchart illustrates the step-by-step selection of patients for our LT study cohort. Starting with 43,249 patients (710,806 notes) mentioning liver transplant, we applied multiple exclusion criteria based on social work interaction, demographic factors, age, encounter frequency, presence of psychosocial evaluations, and availability of clinical data. The final cohort comprised 3,704 patients with complete psychosocial evaluations, demographic factors, and clinical data, including MELD scores within 31 days of evaluation. Proportions for listing and transplant receipt at UCSF are provided for the cohort with psychosocial evaluation notes.}
        \label{fig:cohort_diagram}
\end{figure*}

\begin{table}[htbp]
\centering
\small
\setlength{\tabcolsep}{6pt}
\caption{Demographic Characteristics and Listing Outcomes for Evaluated Patients}
\label{table:demographics_by_list_LT}
\begin{tabular}{lccc}
\hline
 & \multicolumn{3}{c}{Listed} \\[0.5ex]
\cline{2-4}
\multicolumn{1}{l}{Characteristic} & No & Yes & P-value \\
 & (n=1066) & (n=3265) & \\[0.5ex]
\hline
Race-Ethnicity & & & \\
\hspace{1em}Asian & 100 (9.4) & 414 (12.7) & \\
\hspace{1em}Black/African American & 48 (4.5) & 137 (4.2) & \\
\hspace{1em}Hispanic/Latino & 313 (29.4) & 1023 (31.3) & $<$0.001 \\
\hspace{1em}Indigenous/Pacific & 17 (1.6) & 66 (2.0) & \\
\hspace{1em}Non-Hispanic White & 428 (40.2) & 1361 (41.7) & \\
\hspace{1em}Other & 63 (5.9) & 156 (4.8) & \\
\hspace{1em}Unknown/Declined & 97 (9.1) & 108 (3.3) & \\[0.5ex]
Sex & & & \\
\hspace{1em}Female & 469 (44.0) & 1223 (37.5) & $<$0.001 \\
\hspace{1em}Male & 597 (56.0) & 2042 (62.5) & \\
\hline
\multicolumn{4}{l}{\footnotesize Note: Data presented as n (\%). P-values calculated using chi-squared tests.} \\
\end{tabular}
\end{table}

\begin{figure}[htbp]
\centering
\includegraphics[width=1.0\textwidth]{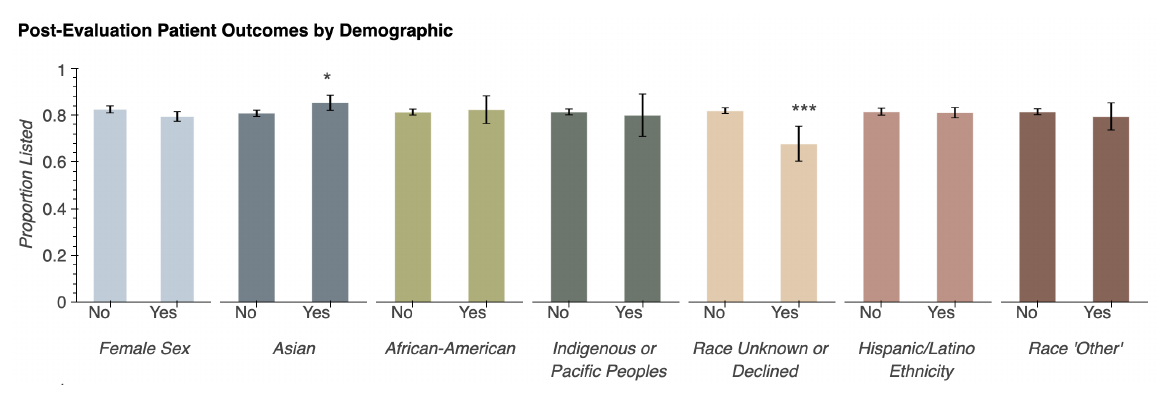}
\caption{\textbf{Proportions of Patients Listed by Demographic Subgroup}. This figure displays the proportions of evaluated patients with full clinical data (n=3,704) listed for liver transplant across demographic subgroups. Error bars represent 95\% confidence intervals. Statistical significance, determined using proportions z-tests, is denoted by asterisks (* p$<$0.05, ** p$<$0.01, *** p$<$0.001).}
\label{fig:simple_dem_comp}
\end{figure}

\begin{figure}[htbp]
\centering
\includegraphics[width=1.0\textwidth]{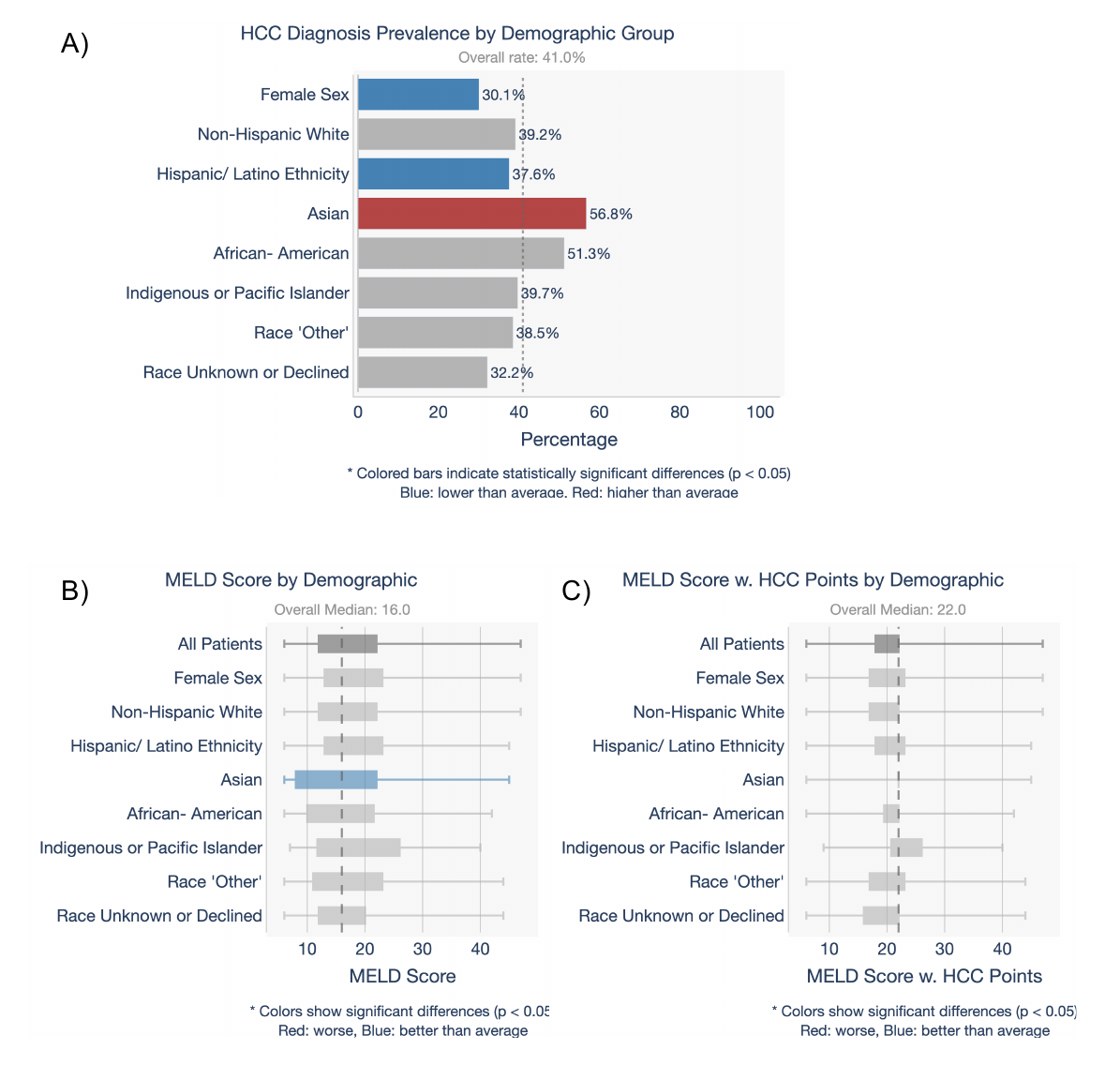}
\caption{\textbf{Demographic distribution of liver disease severity metrics.} a) Hepatocellular carcinoma (HCC) prevalence across demographic groups. Colored bars indicate statistically significant differences from overall cohort mean (Fisher's exact test, p$<$0.05, FDR-corrected); blue indicates lower prevalence, red indicates higher prevalence. b) Distribution of laboratory MELD scores by demographic group. c) Distribution of MELD scores including HCC exception points (standardized to minimum score of 22 for HCC patients) by demographic group. For (b) and (c), box plots show median, interquartile range, and whiskers at 1.5× IQR; significance assessed by Kruskal-Wallis H test.}
\label{fig:liver_disease_comparison}
\end{figure}
\clearpage

\subsection{LLM-based Extraction from Clinical Notes: Detailed Methods}
\label{apd:appendix_llm_methods}

We present details on our LLM-based information extraction strategy including the prompt and the expert-informed questions and categories provided to the LLM in Table \ref{tab:llm_questions_categories}.

\subsubsection{LLM Prompt}
Prompt tuning with multiple system and task prompts was carried out on a small selection (n=20) of expert-labeled notes disjoint from the evaluation set. Following this process, the following prompt was used to instruct the GPT-4-Turbo-128k model for information extraction from clinical notes:

\begin{quote}
\small
\emph{Assume the role of an expert medical professional. Your task is to extract and interpret vital information from clinical notes accurately. You will be provided with a clinical note, enclosed by triple backticks, and a set of questions. For each question related to the notes, choose the most accurate category (label) based on the evidence in the note. Format your analysis as JSON with 'Question Number' and 'Label'. If no evidence supports a category, return {``Question Number'': [number], ``Label'': ``No evidence''}. Ensure precision in label identification and documentation. If multiple categories apply, select the most relevant one and justify your choice briefly. For the last two questions where no category choices are specified, answer each question in 50 words or less based on the note content and return that answer as the ``Label''.}
\end{quote}

\begin{figure*}[htbp]
    \centering
    \caption{\textbf{Confusion matrices for LLM-derived information (Part 1)}}
    \includegraphics[width=0.9\textwidth]{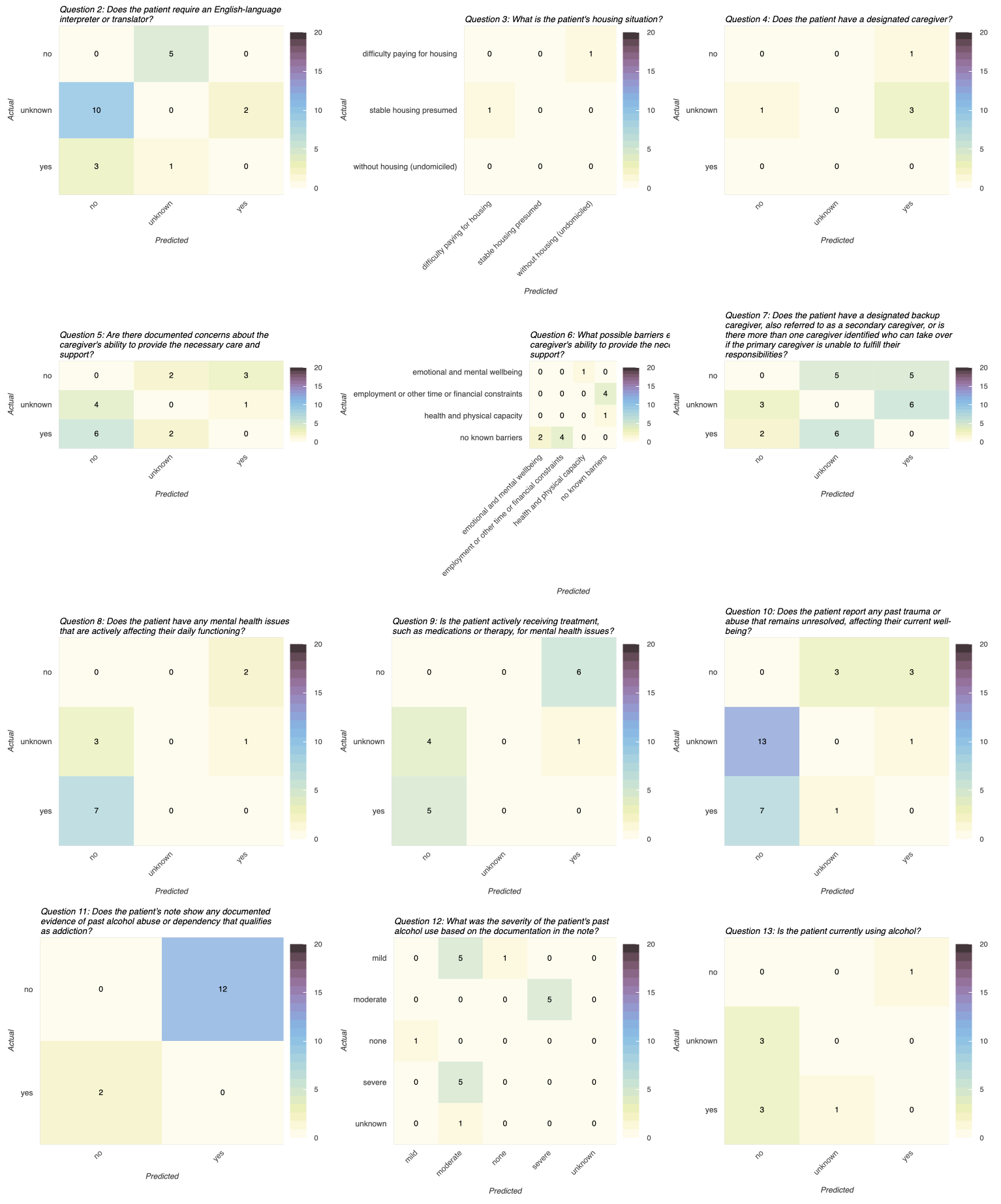}
        \label{fig:cm_llm_mismatches_1}
    \end{figure*}

\begin{figure*}[htbp]
    \centering
    \caption{\textbf{Confusion matrices for LLM-derived information (Part 2)}}
    \includegraphics[width=0.9\textwidth]{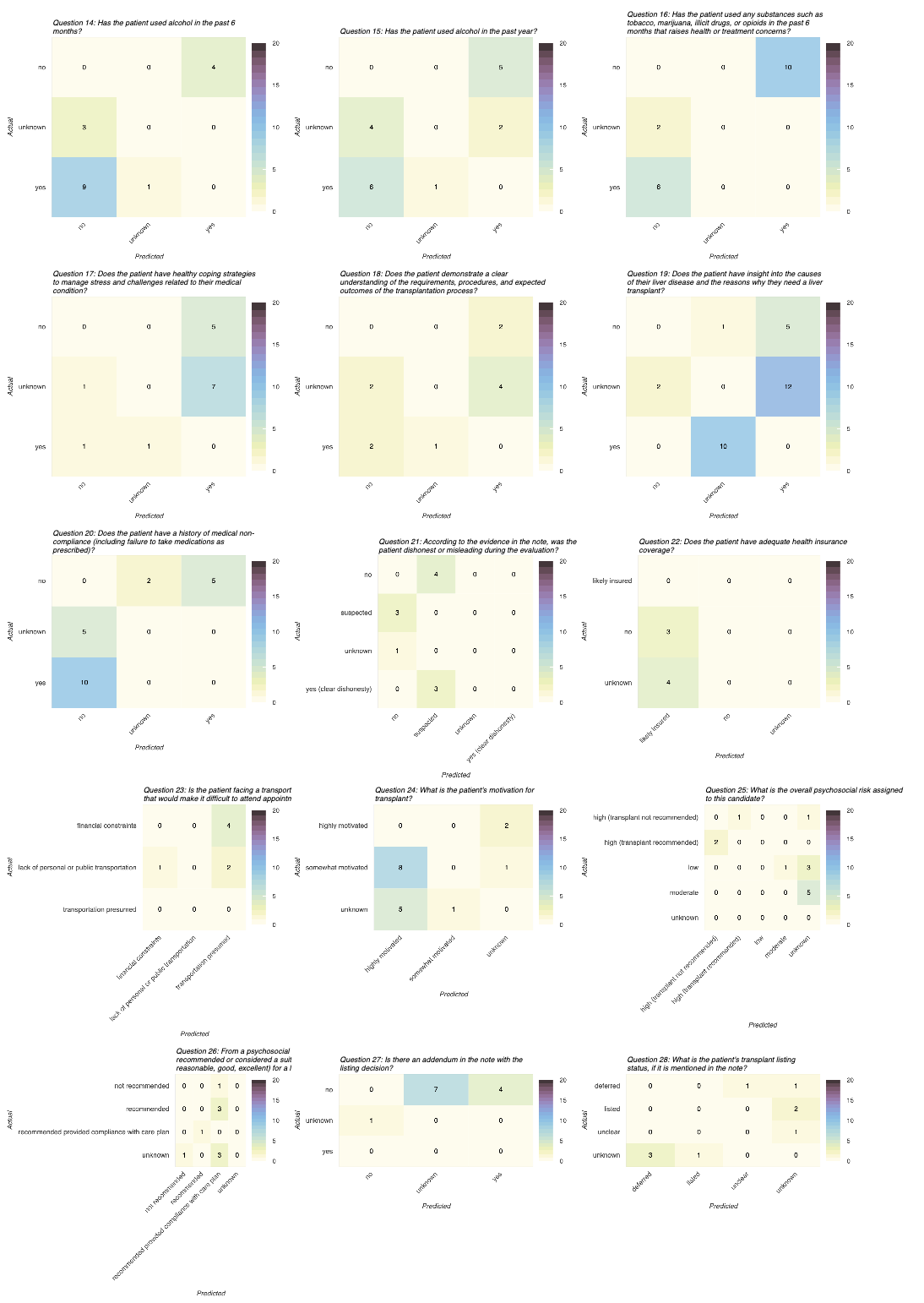}
        \label{fig:cm_llm_mismatches_2}
    \end{figure*}

Categories returned with ``No Label'', ``NA'', or ``No Evidence'' were mapped to ``Unknown.'' Following the mapping of categories, features were created by one-hot encoding the unique question category pairs for each note. Figures \ref{fig:cm_llm_mismatches_1} and \ref{fig:cm_llm_mismatches_2} present confusion matrices for LLM-derived information. 

\begin{table}[htbp]
\caption{Domain Expert-Informed Annotation Questions for LLM Query of Psychosocial Evaluation Notes}
\small
\begin{tabular}{p{0.02\textwidth}p{0.95\textwidth}}
\toprule
\# & Question [Categories] \\
\midrule
1 & Does the note specifically provide a psychosocial evaluation addressing the patient's suitability for a liver transplant? [\emph{Yes, No, Unknown}] \\
\hline
2 & Does the patient require an English-language interpreter or translator? [\emph{Yes, No, Unknown}] \\
\hline
3 & What is the patient's housing situation? [\emph{Stable Housing, Difficulty Paying for Housing, Without Housing (Undomiciled), Unknown}] \\
\hline
4 & Does the patient have a designated caregiver? [\emph{Yes, No, Unknown}] \\
\hline
5 & Are there documented concerns about the caregiver's ability to provide the necessary care and support? [\emph{Yes, No, Unknown}] \\
\hline
6 & What possible barriers exist regarding the caregiver's ability to provide the necessary care and support? [\emph{Health and Physical Capacity, Emotional and Mental Wellbeing, Employment or other Time or Financial Constraints, No Known Barriers, Unknown}] \\
\hline
7 & Does the patient have a designated backup caregiver, also referred to as a secondary caregiver, or is there more than one caregiver identified who can take over if the primary caregiver is unable to fulfill their responsibilities? [\emph{Yes, No, Unknown}] \\
\hline
8 & Does the patient have any mental health issues that are actively affecting their daily functioning? [\emph{Yes, No, Unknown}] \\
\hline
9 & Is the patient actively receiving treatment, such as medications or therapy, for mental health issues? [\emph{Yes, No, Unknown}] \\
\hline
10 & Does the patient report any past trauma or abuse that remains unresolved, affecting their current well-being? [\emph{Yes, No, Unknown}] \\
\hline
11 & Does the patient's note show any documented evidence of past alcohol abuse or dependency that qualifies as addiction? [\emph{Yes, No, Unknown}] \\
\hline
12 & What was the severity of the patient's past alcohol use based on the documentation in the note? [\emph{None, Mild, Moderate, Severe, Unknown}] \\
\hline
13 & Is the patient currently using alcohol? [\emph{Yes, No, Unknown}] \\
\hline
14 & Has the patient used alcohol in the past 6 months? [\emph{Yes, No, Unknown}] \\
\hline
15 & Has the patient used alcohol in the past year? [\emph{Yes, No, Unknown}] \\
\hline
16 & Has the patient used any substances such as tobacco, marijuana, illicit drugs, or opioids in the past 6 months that raises health or treatment concerns? [\emph{Yes, No, Unknown}] \\
\hline
17 & Does the patient have healthy coping strategies to manage stress and challenges related to their medical condition? [\emph{Yes, No, Unknown}] \\
\hline
18 & Does the patient demonstrate a clear understanding of the requirements, procedures, and expected outcomes of the transplantation process? [\emph{Yes, No, Unknown}] \\
\hline
19 & Does the patient have insight into the causes of their liver disease and the reasons why they need a liver transplant? [\emph{Yes, No, Unknown}] \\
\hline
20 & Does the patient have a history of medical non-compliance (including failure to take medications as prescribed)? [\emph{Yes, No, Unknown}] \\
\hline
21 & According to the evidence in the note, was the patient dishonest or misleading during the evaluation? [\emph{Yes, Suspected, No, Unknown}] \\
\hline
22 & Does the patient have adequate health insurance coverage? [\emph{Yes, No, Pending Confirmation, Unknown}] \\
\hline
23 & Is the patient facing a transportation issue that would make it difficult to attend appointments? [\emph{Distance/Travel Time, Lack of Personal or Public Transportation, Financial Constraints, No Transportation Issues, Unknown}] \\
\hline
24 & What is the patient's motivation for transplant? [\emph{Highly Motivated, Somewhat Motivated, Not Motivated, Unknown}] \\
\hline
25 & What is the overall psychosocial risk assigned to this candidate? [\emph{Low, Moderate, High (Transplant Recommended), High (Transplant Not Recommended), Unknown}] \\
\hline
26 & From a psychosocial perspective, is the patient recommended or considered a suitable candidate (e.g., reasonable, good, excellent) for a liver transplant? [\emph{Recommended, Recommended Provided Compliance with Care Plan, Not Recommended, Unknown}] \\
\hline
27 & Is there an addendum in the note with the listing decision? [\emph{Yes, No, Unknown}] \\
\hline
28 & What is the patient's transplant listing status, if it is mentioned in the note? [\emph{Listed, Deferred, Declined/Denied, Status 1A, Temporarily Unfit, Unclear, Unknown}] \\
\hline
29 & What specific risk factors or concerns have been reported that could impact the patient's suitability and fitness for a liver transplant? [\emph{Open-ended}] \\
\hline
30 & What specific protective factors have been reported that enhance the patient's suitability and fitness for a liver transplant? [\emph{Open-ended}] \\
\bottomrule
\end{tabular}
\label{tab:llm_questions_categories}
\end{table}

\clearpage

\subsection{SDOH Factor Prevalence Expanded Results}

\begin{figure}
    \centering
    \includegraphics[width=0.95\linewidth]{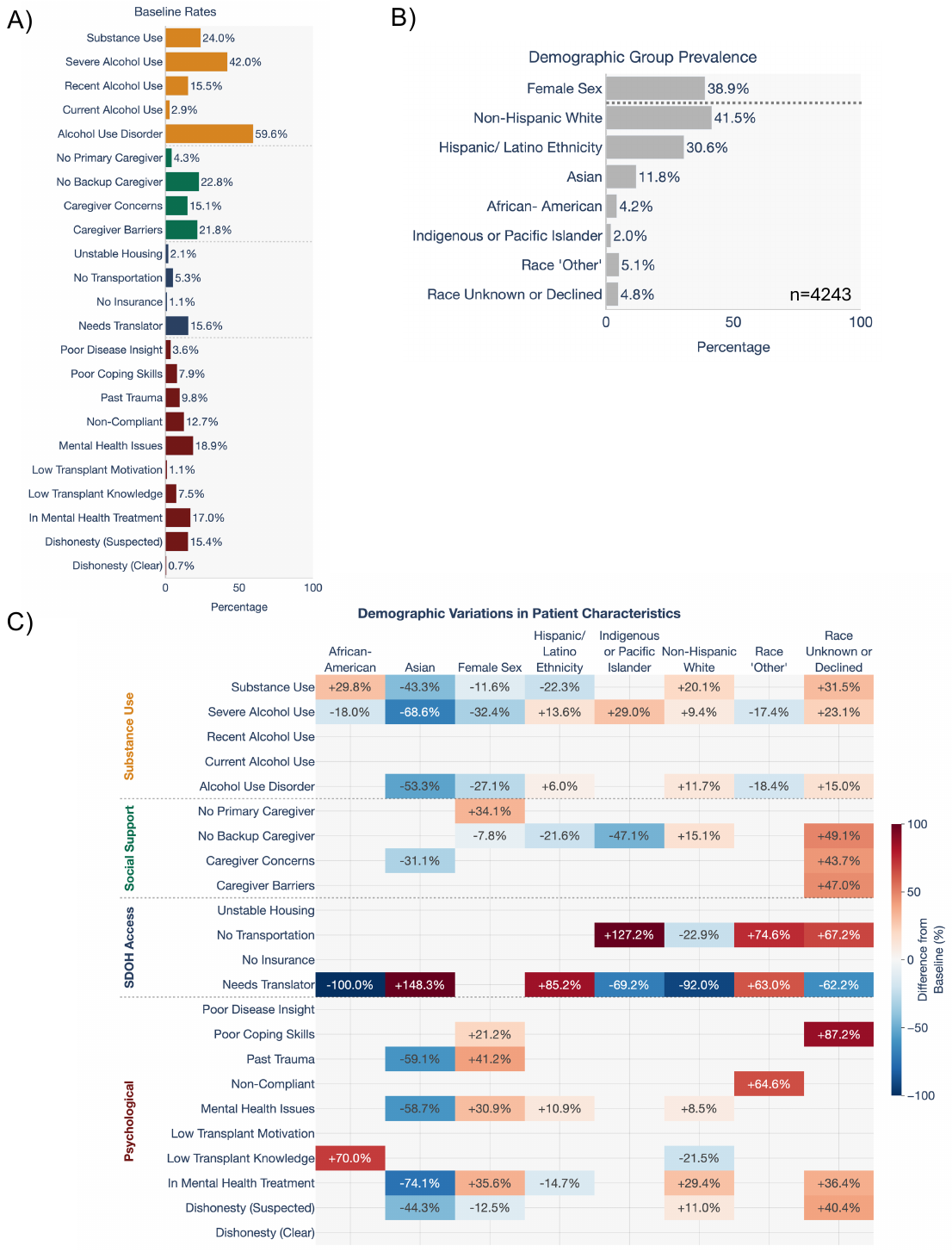}
    \caption{\textbf{Analysis of demographic disparities in liver transplant listing rates across all patients (n=4243)}. a) Baseline prevalence rates for psychosocial and substance use factors identified in clinical notes. b) Demographic composition of the study cohort (n=4,243). c) Heat map of statistically significant differences in SDOH factor prevalence across patient demographics compared with the cohort average (two-proportion z-tests, p $<$ 0.05, FDR-corrected); blue indicates higher rates, red indicates lower rates, blank cells indicate non-significant differences.}
    \label{fig:dem_var_all_patients}
\end{figure}

\begin{figure}
    \centering
    \includegraphics[width=0.95\linewidth]{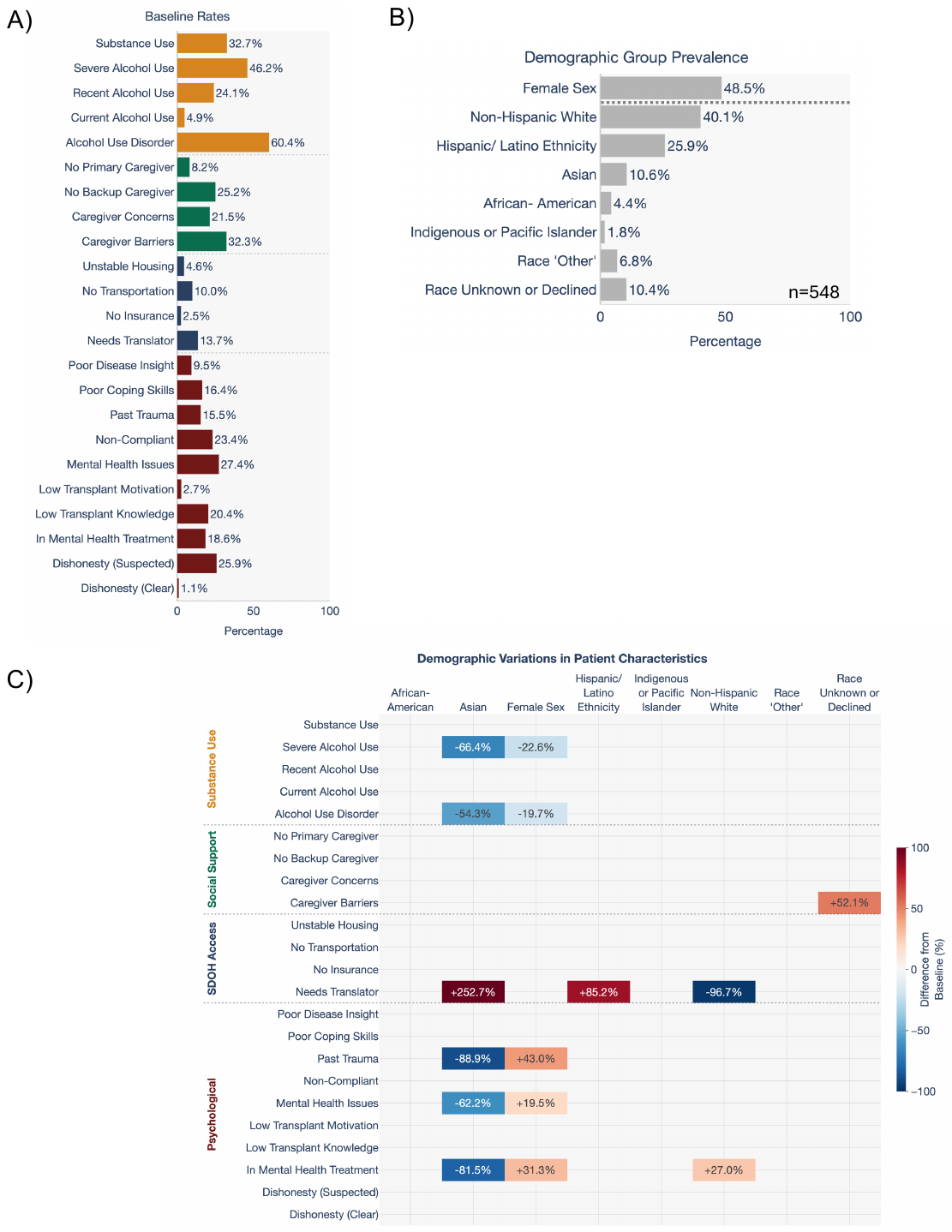}
    \caption{\textbf{Analysis of demographic disparities in liver transplant listing rates across patients missing data (n=548)} a) Baseline prevalence rates for psychosocial and substance use factors identified in clinical notes. b) Demographic composition of the study cohort (n=548). c) Heat map of statistically significant differences in SDOH factor prevalence across patient demographics compared with the cohort average (two-proportion z-tests, p $<$ 0.05, FDR-corrected); blue indicates higher rates, red indicates lower rates, blank cells indicate non-significant differences.}
    \label{fig:dem_var_missing_meta}
\end{figure}

\begin{figure}[htbp]
    \centering
    \begin{subfigure}{1.0\linewidth}
        \centering
        \includegraphics[width=\linewidth, height=0.45\textheight, keepaspectratio]{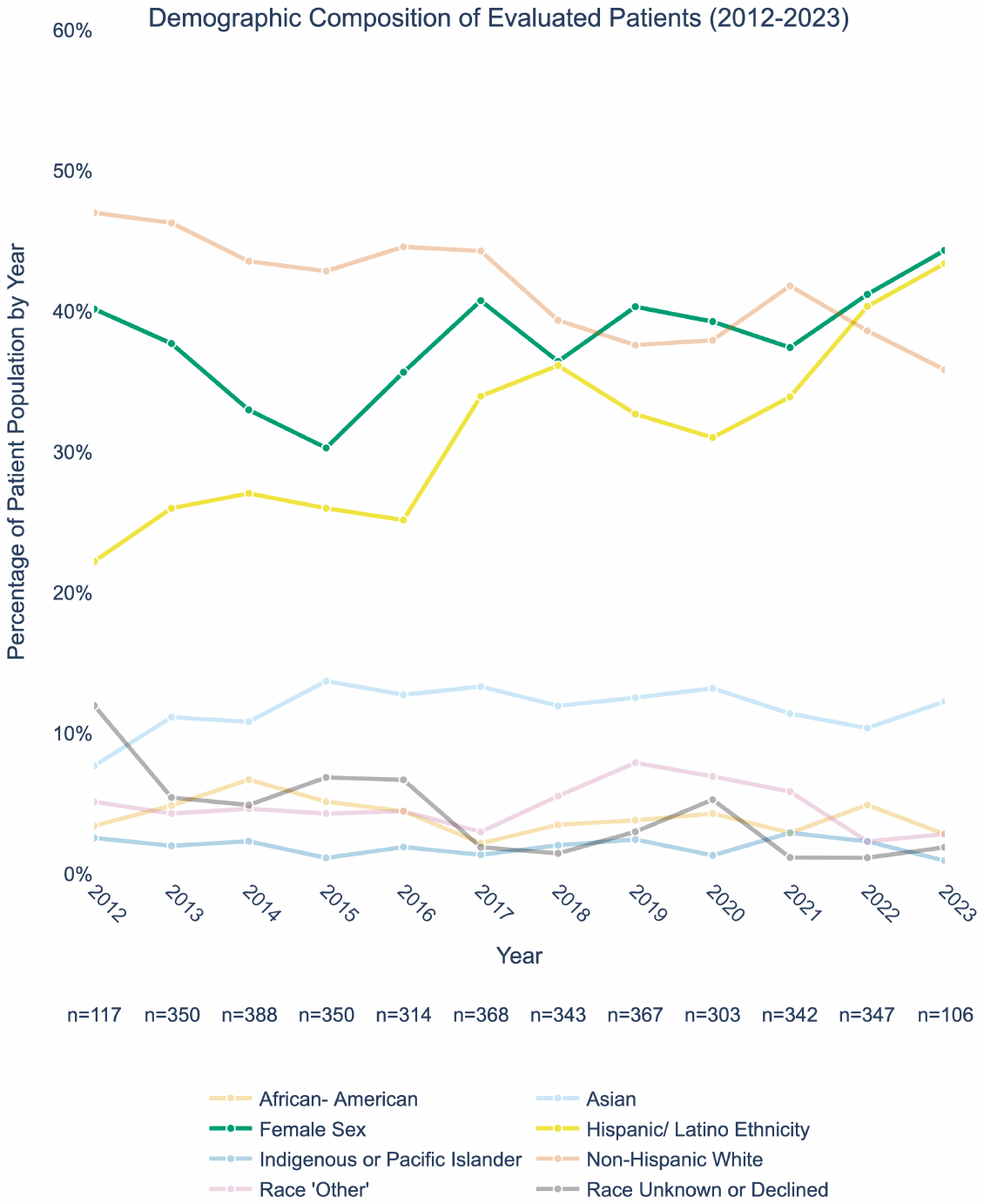}
        \caption{Shifting demographics of patients evaluated for LT (2012-2023)}
        \label{fig:year_dem}
    \end{subfigure}
    
    \begin{subfigure}{1.0\linewidth}
        \centering
        \includegraphics[width=\linewidth, height=0.45\textheight, keepaspectratio]{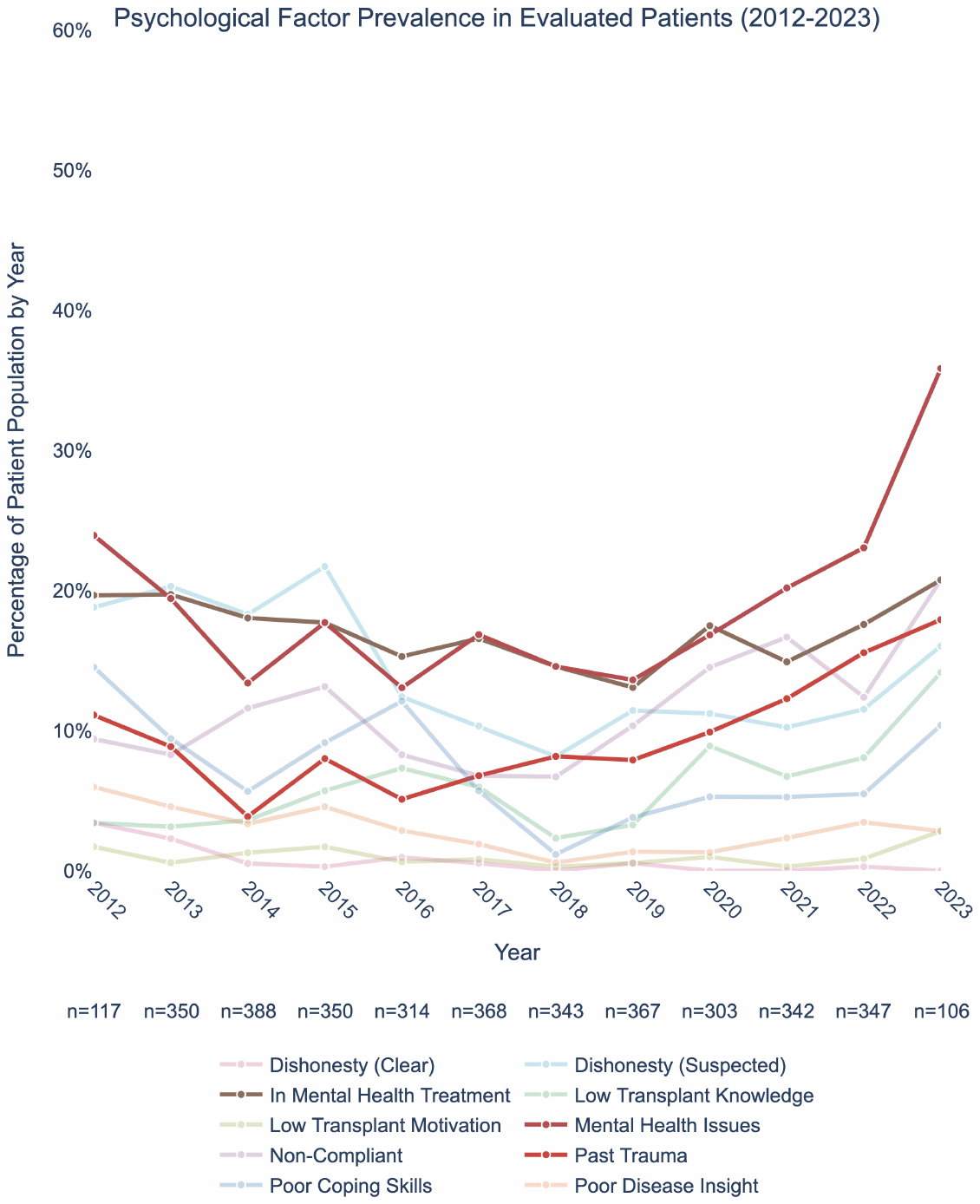}
        \caption{Psychological profile trends in patients evaluated for LT (2012-2023)}
        \label{fig:year_psych}
    \end{subfigure}
    \caption{\textbf{Temporal shifts in patient demographics and psychological factors for transplant evaluations (2012-2023).} (a) Trends in patient demographics over time (b) Trends in prevalence of assessed behavioral and mental health factors over time.}
    \label{fig:lt_trends}
\end{figure}

\begin{figure}[htbp]
    \centering
    \begin{subfigure}{1.0\linewidth}
        \centering
        \includegraphics[width=\linewidth, height=0.45\textheight, keepaspectratio]{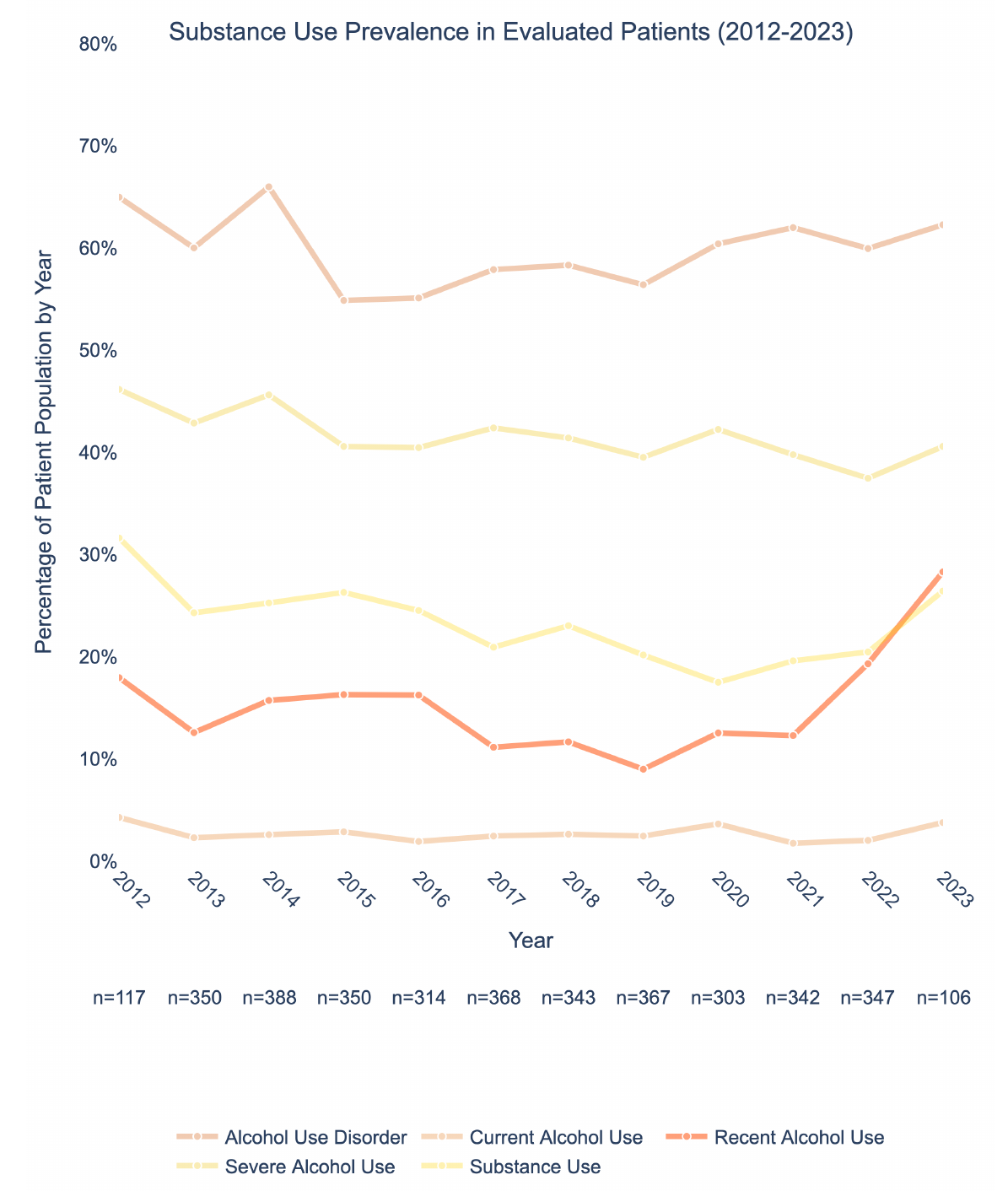}
        \caption{Shifting substance us of patients evaluated for LT (2012-2023)}
        \label{fig:year_substance}
    \end{subfigure}
    
    \begin{subfigure}{1.0\linewidth}
        \centering
        \includegraphics[width=\linewidth, height=0.45\textheight, keepaspectratio]{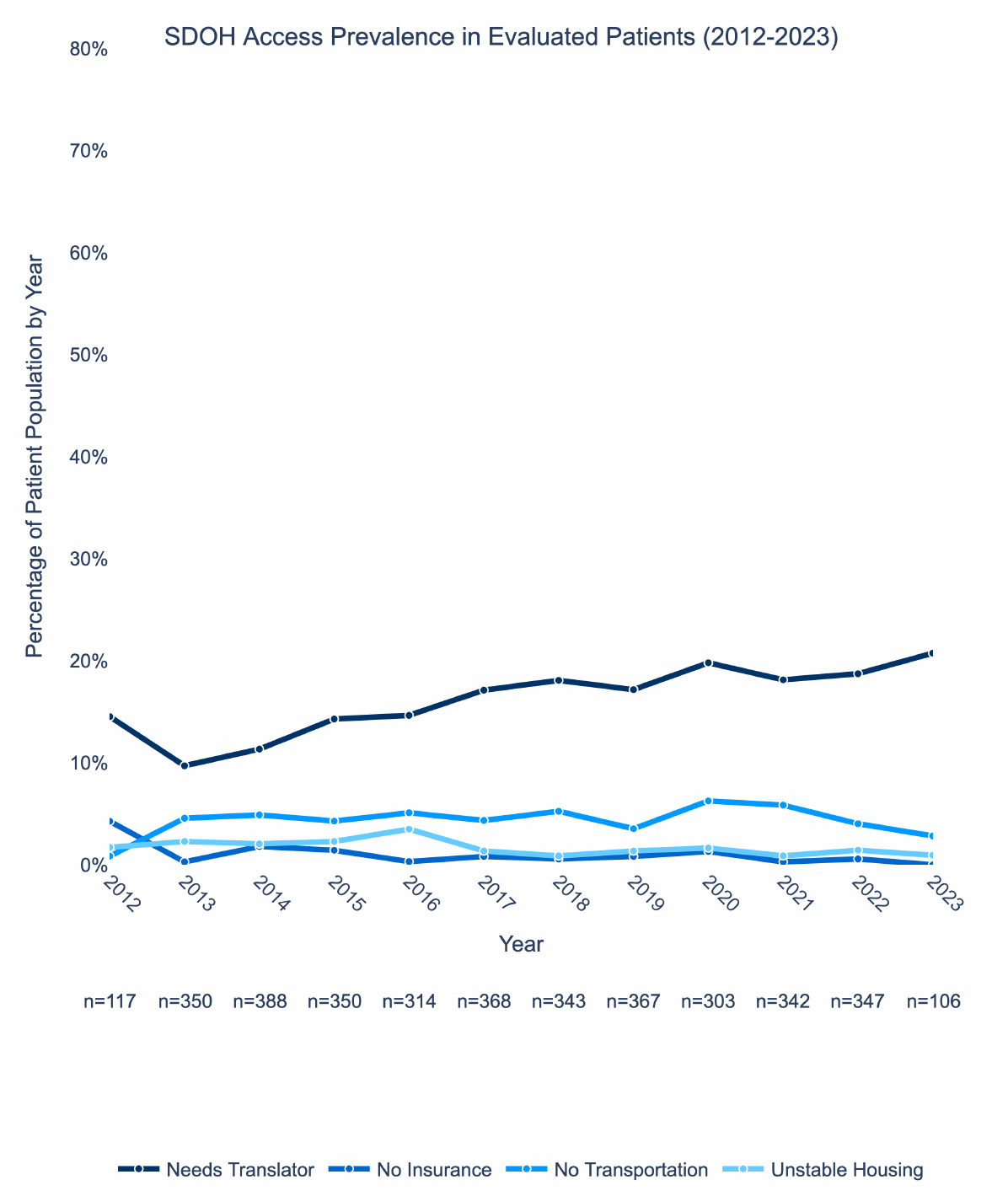}
        \caption{Access-relevant trends in patients evaluated for LT (2012-2023)}
        \label{fig:year_access}
    \end{subfigure}
    \caption{\textbf{Temporal shifts in patient substance use and access-related SDOH factors for transplant evaluations (2012-2023).} (a) Trends in prevalence of patient substance use over time (b) Trends in prevalence of assessed access factors over time.}
    \label{fig:lt_trends_2}
\end{figure}

\begin{figure}[htbp]
    \centering
    \includegraphics[width=0.9\textwidth]{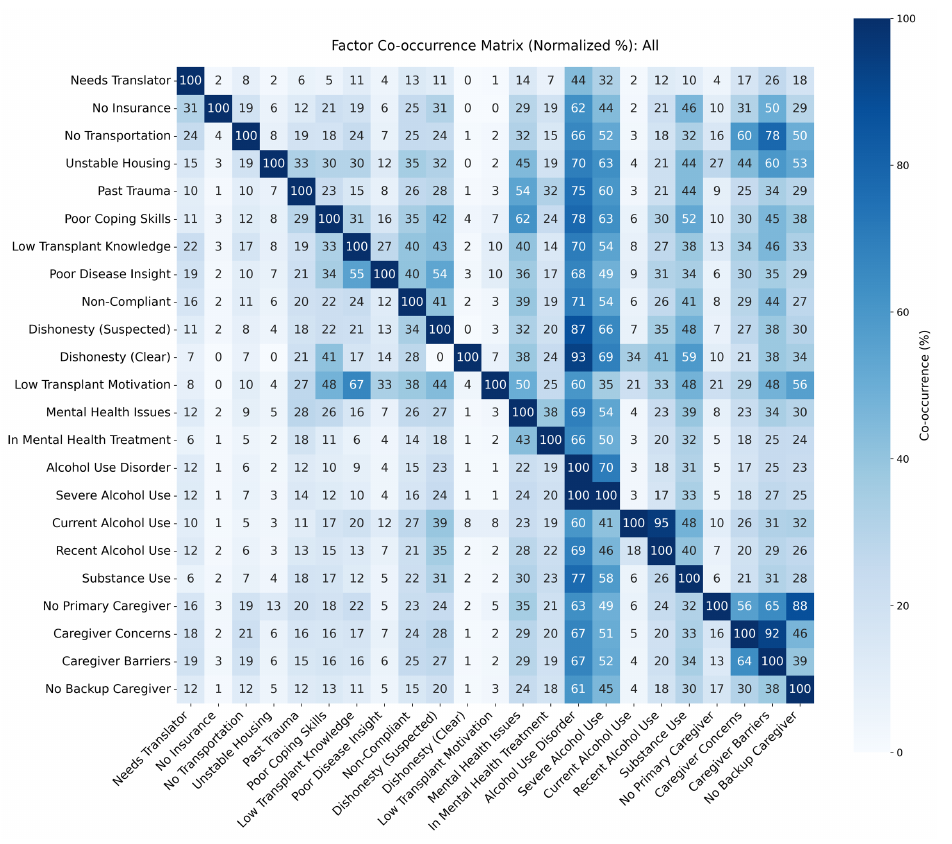}
    \caption{\textbf{Adverse SDOH co-occurrence matrix.} Heat map showing normalized pairwise co-occurrence of SDOH factors across patients; darker blue indicates higher rates.}
        \label{fig:sdoh_cooccur}
    \end{figure}

\begin{figure}[htbp]
    \centering
    \includegraphics[width=0.9\textwidth]{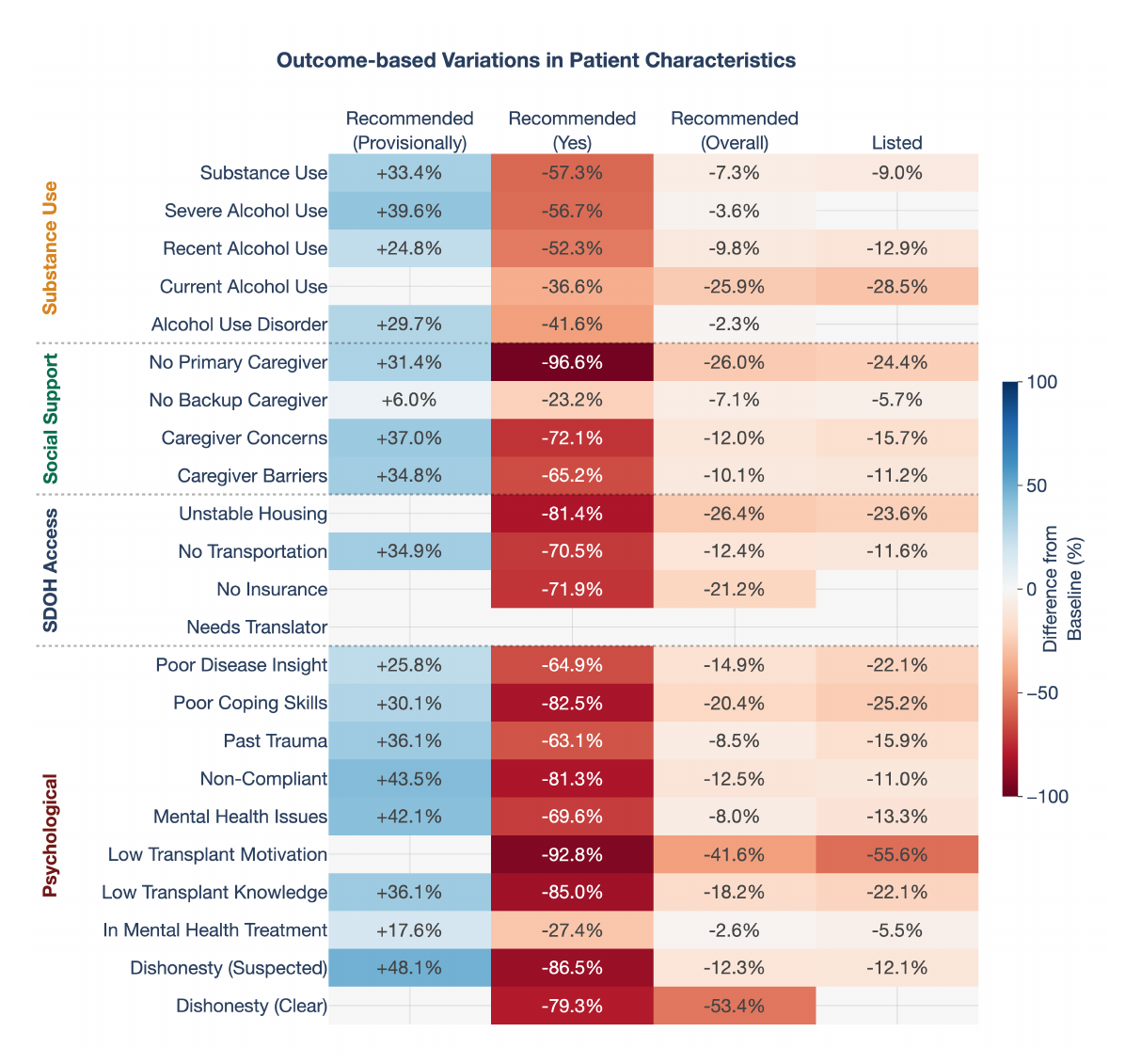}
    \caption{\textbf{SDOH and their relationship to expanded psychosocial recommendation designations and listing.} Heat map showing only statistically significant differences in SDOH factor prevalence between patients who did versus did not achieve each outcome (two-proportion z-tests, p $<$ 0.05, FDR-corrected); blue indicates higher rates, red indicates lower rates, blank cells indicate non-significant differences.}
        \label{fig:sdoh_ext_rec}
    \end{figure}

\clearpage

\subsection{Extended Data: Enhanced Accuracy and Interpretability of LLM-Derived SDOH Features}\label{secA1} 

\subsubsection{Text Feature Processing}
BOW features were derived from raw note text that was cleaned and preprocessed with the NLTK Python package \citep{bird2009natural}. Removal of standard stop words and lemmatization with the WordNetLemmatizer was applied prior to count vectorization with the scikit-learn CountVectorizer with an ngram range of 1-2. We filter for terms that appeared in more than five notes, but fewer than 80\% of the notes, with a maximum vocabulary size of 10,000 terms. We standardize data with StandardScaler and the top 100 features based on a chi-squared test with the outcome were chosen with the scikit-learn SelectKBest and chi2 implementations. For the cTAKES concepts based model, concepts that appear in $>$0.1\% of notes were selected, counted, and the prevalence was also standardized with StandardScaler.

\subsubsection{Detailed Results}

\begin{figure}[htbp]
    \centering
    \includegraphics[width=0.7\textwidth]{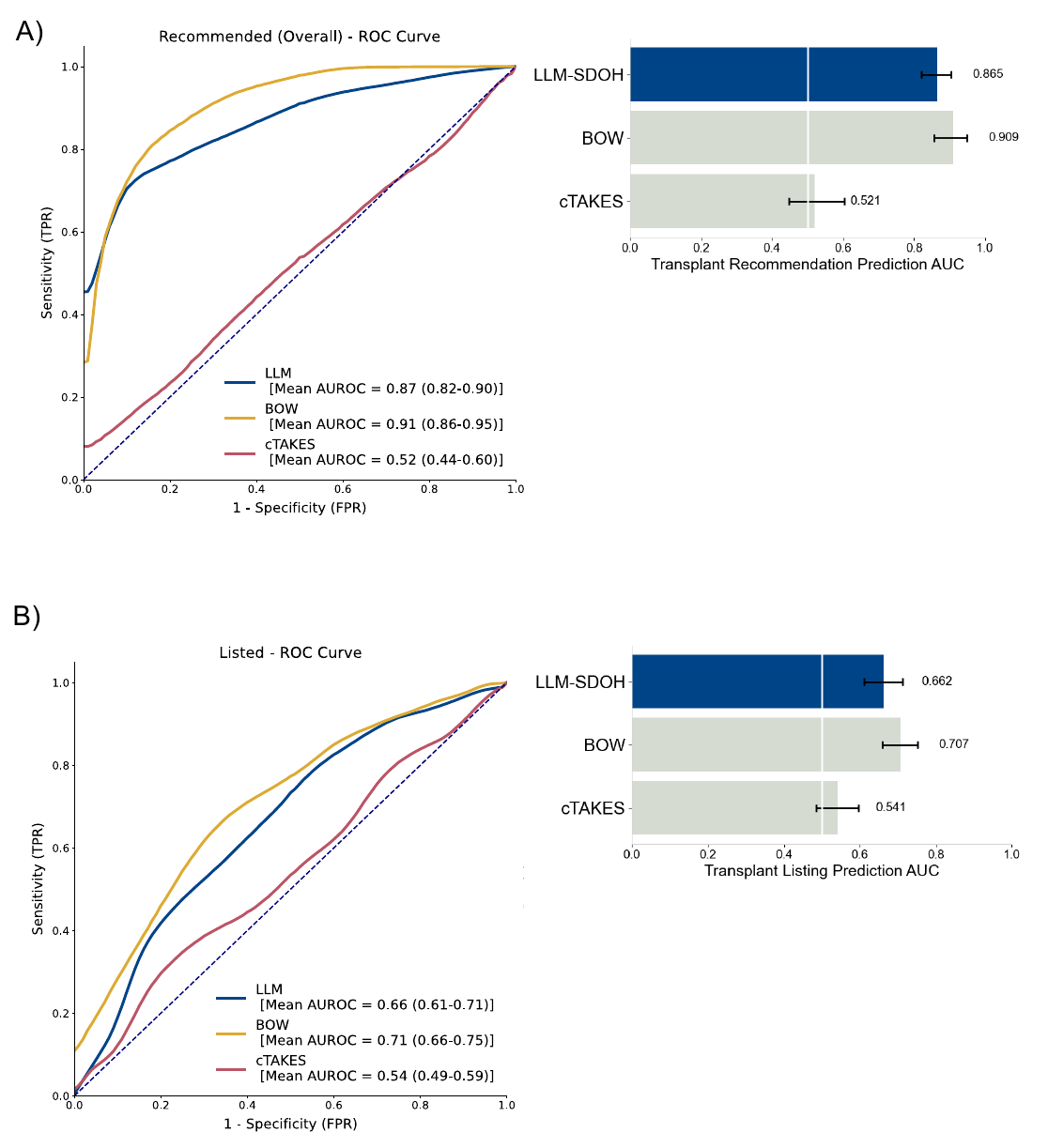}
    \caption{\textbf{Predictive Capacity of Text-based Features in LT Process.} a) AUROC curves for XGBoost model performance using LLM-derived, BOW, and cTAKES features to predict psychosocial recommendation (left). Comparison of average AUROC (w. 95\% CI) across the feature sets. Feature sets including LLM-derived features highlighted in blue (right). b) Same as (a) but for models predicting transplant listing}
        \label{fig:rec_curves_text_only}
    \end{figure}

Model performance metrics for the XGBoost models based on BOW, cTAKES, or LLM-derived features across the two binary outcomes (recommendation, listing) are summarized in Table \ref{tab:performance_text_models} with ROC curves shown in Figure \ref{fig:rec_curves_text_only}.

\begin{table}[htbp]
\centering
\caption{Performance of Text-Based XGBoost Models Across Transplant Evaluation Decisions}
\label{tab:performance_text_models}
\small
\setlength{\tabcolsep}{6pt}
\begin{tabular}{@{}p{1.8cm}lccc@{}}
\toprule
Outcome & Model & AUROC & Sensitivity & Specificity \\
\midrule
\multirow{3}{*}{Rec. (Overall)} 
& BOW & 0.91 (0.89--0.93) & 0.92 (0.90--0.94) & 0.69 (0.57--0.82) \\
& LLM & 0.87 (0.84--0.89) & 0.80 (0.77--0.83) & 0.76 (0.64--0.88) \\
& cTAKES & 0.52 (0.49--0.56) & 0.50 (0.46--0.54) & 0.54 (0.40--0.68) \\
\midrule
\multirow{3}{*}{Listed} 
& BOW & 0.71 (0.68--0.74) & 0.69 (0.65--0.73) & 0.65 (0.57--0.73) \\
& LLM & 0.66 (0.63--0.70) & 0.72 (0.69--0.76) & 0.50 (0.42--0.58) \\
& cTAKES & 0.54 (0.50--0.58) & 0.59 (0.55--0.63) & 0.44 (0.35--0.52) \\
\bottomrule
\end{tabular}
\begin{tablenotes}[center]
\small
\item Note: Values shown as mean (95\% confidence interval); 0.5 threshold used for sensitivity/specificity calculations. BOW = Bag of Words; cTAKES = clinical Text Analysis and Knowledge Extraction System.
\end{tablenotes}
\end{table}

\subsubsection{Feature Interpretability Comparison with BOW and cTAKES Features}
\label{sec:interpretability}
Here we elaborate on the specific advantages and limitations of the LLM-derived features derived from our approach relevant to their interpretability and utility, with a focus on BOW versus LLM-derived features. \newline

\begin{figure}[htbp]
    \centering
    \includegraphics[width=1.0\textwidth]{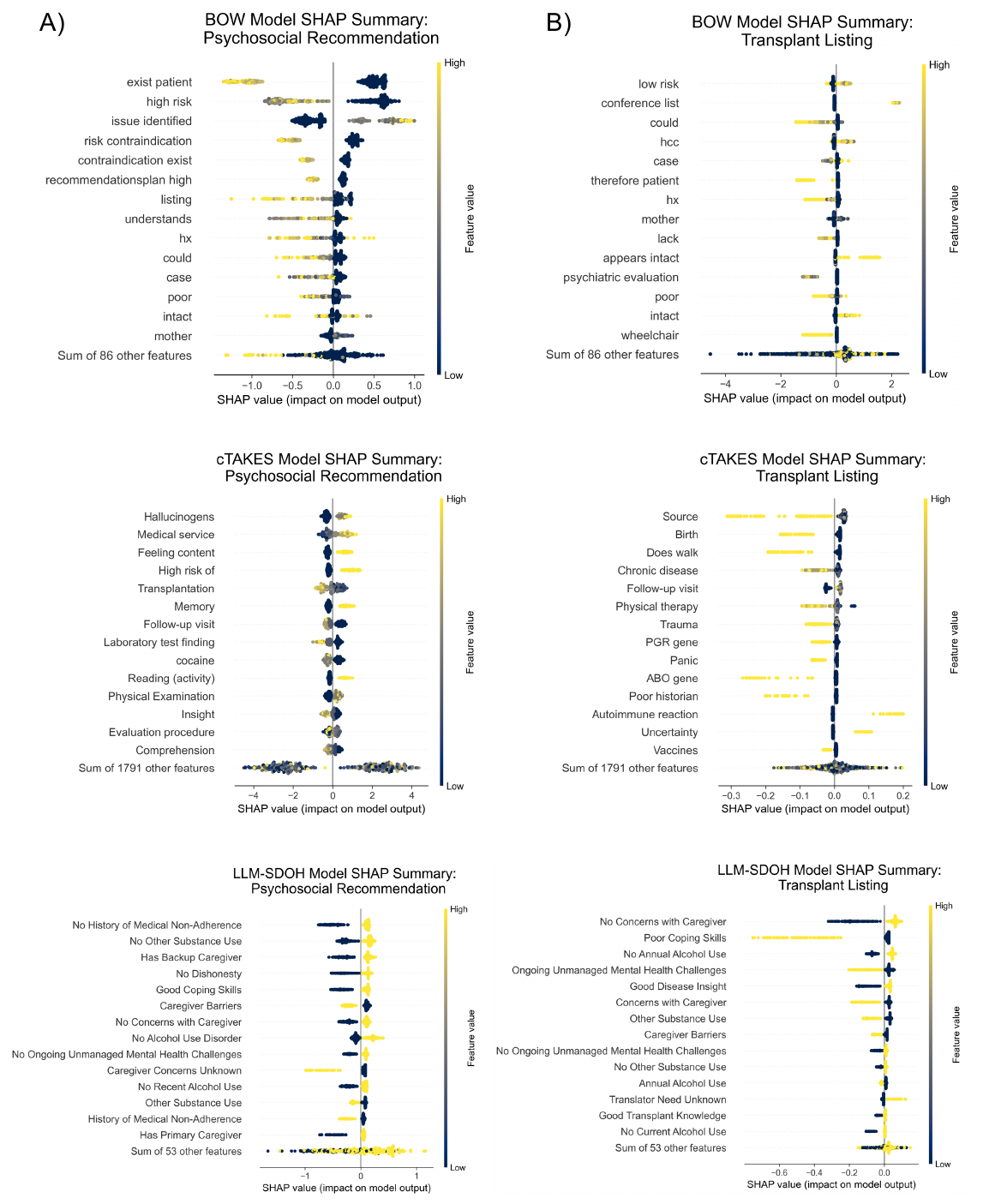}
    \caption{\textbf{SHAP Value Summaries for Models Solely Based on Text Features.} a) SHAP values for XGBoost models predicting psychosocial recommendation based on Bag-of-Words (BOW) (top), clinical Text Analysis and Knowledge Extraction System (c-TAKES) (middle), and LLM-derived features (bottom).  b) The same as (a) but for models predicting transplant listing. Features ranked by model importance; blue indicates lower feature values, yellow indicates higher feature values.}
        \label{fig:shap_text_only}
    \end{figure}

While BOW and LLM-derived features showed similar predictive performance, LLM-derived features offer superior interpretability and utility in evaluating LT decision processes. This is evident from the SHAP values for BOW, cTAKES, and LLM-derived models (Figure \ref{fig:shap_text_only}). LLM-derived features have several advantages over BOW features. They maintain temporal, relational, and negation context, which BOW features often lose. For example, an LLM can accurately answer ``Does the patient have any mental health issues that are actively affecting their daily functioning?'', while a BOW approach might only capture the presence of the term ``mental health issues'' without context. LLM-derived features are directly interpretable and designed to capture factors relevant to LT decision-making. For example, the question ``What is the severity of the patient's past alcohol use based on the documentation in the note?'' provides clear, categorized information (None, Mild, Moderate, Severe, Unknown) that is directly relevant to LT evaluation. In contrast, BOW features may include ambiguous terms like ``poor'', ``lack'', ``could'', or ``intact'' without clear context or relevance. The LLM approach allows for specification of timing and valence, which is typically unclear in BOW features. For instance, the question ``Has the patient used alcohol in the past 6 months?'' provides a clear timeframe, while a BOW feature might only indicate the presence of alcohol-related terms without temporal context. Another advantage of LLM-derived features is that they lead to an automatically standardized feature set across all notes, whereas BOW features vary based on note content. This standardization allows for more consistent analysis across patients, with each note being evaluated on the same set of expert-informed questions. LLM-derived features also capture information at a relevant granularity for LT decision-making. For example, the question ``What is the patient's housing situation?'' provides categories (Stable Housing, Difficulty Paying for Housing, Without Housing (Undomiciled), Unknown) that are directly relevant to assessing a patient's stability and support system. BOW features, in contrast, might include overly specific details that are less directly relevant to the overall assessment.

However, LLM-derived features have limitations, including possible hallucinations. The way we employ the LLM to essentially ``survey'' a note introduces concerns similar to survey designs, such as the limitation of only obtaining information we think to ask about and the impact of question wording on extracted information. There's also the possibility that information to answer a particular question may not be mentioned in the note, although this is addressed by including an ``Unknown'' category in the LLM's response options for each question. The BOW approach, while less targeted, may capture unexpected relevant information. Despite these trade-offs, the interpretability and relevance of LLM-derived features make them particularly useful for analyzing decision points and outcomes in the LT process. In addition, it is worth noting that BOW features may inadvertently include terms directly related to outcomes (e.g., ``recommendation'', ``listing''), potentially leading to label leakage. LLM-derived features can be designed avoid this issue by focusing on patient characteristics rather than process outcomes. They can also be designed to accurately extract such labels that are absent in the structured data-as we do with the psychosocial recommendation-so they can be used (with some caution) as targets for prediction tasks. Overall, the use of LLM-derived SDOH features leads to models with similar predictive power as text-based models. However, it's important to note that some of this predictive power in BOW models may come from terms that directly mention recommendation and risk, which are difficult to fully remove from the notes. 

\end{document}